\documentclass[aps, twocolumn,superscriptaddress, 
amsmath,amssymb,reprint,numbers,noeprint]{revtex4-1}
\usepackage{graphicx} 
\usepackage{appendix}
\usepackage{xcolor}
\begin{document}

\title{Direct measure of DNA bending by quantum magnetic imaging of a nano-mechanical torque-balance}
\author{Zeeshawn Kazi}
\email{zeeshawn@princeton.edu}
\altaffiliation{Present address: Department of Electrical and Computer Engineering, Princeton University, Princeton, NJ 08540, USA}
\affiliation{Department of Physics, University of Washington, Seattle, WA 98105, USA}%
\author{Isaac M. Shelby}%
\affiliation{Department of Physics, University of Washington, Seattle, WA 98105, USA}%
\author{Ruhee Nirodi}
\affiliation{Department of Physics, University of Washington, Seattle, WA 98105, USA}%
\author{Joseph Turnbull}
\affiliation{Department of Physics, University of Washington, Seattle, WA 98105, USA}%
\author{Hideyuki Watanabe}
\affiliation{National Institute of Advanced Industrial Science and Technology,
Tsukuba Central 2, 1-1-1 Umezono, Tsukuba, Ibaraki 305-8568, Japan}
\author{Kohei M. Itoh}
\affiliation{School of Fundamental Science and Technology, Keio University, 3-14-1 Hiyoshi, Kohoku-ku, Yokohama 223-8522, Japan}
\author{Paul A. Wiggins}
\affiliation{Department of Physics, University of Washington, Seattle, WA 98105, USA}%
\affiliation{Department of Bioengineering, University of Washington, Seattle, WA 98105, USA}
\author{Kai-Mei C. Fu}
\affiliation{Department of Physics, University of Washington, Seattle, WA 98105, USA}%
\affiliation{Department of Electrical and Computer Engineering, University of Washington, Seattle, WA 98105, USA}
\affiliation{Physical Sciences Division, Pacific Northwest National Laboratory, Richland, WA 99354, USA}

\begin{abstract}
    DNA flexibility is a key determinant of biological function, from nucleosome positioning to transcriptional regulation, motivating a direct measurement of the bend-torque response of individual DNA molecules. In this work, DNA bending is detected using a nano-mechanical torque balance formed by tethering a ferromagnetic nanoparticle probe by an individual DNA molecule to a diamond magnetic field imager. The torque exerted by the DNA in response to bending caused by an applied magnetic torque is measured using wide-field imaging of quantum defects near the surface of the diamond. Qualitative measurements of differences in DNA bio-mechanical binding configuration are demonstrated, and as a proof-of-principle, a quantitative measurement of the bend response is made for individual DNA molecules. This quantum-enabled measurement approach could be applied to characterize the bend response of biophysically relevant short DNA molecules as well as the sequence dependence of DNA bending energy.
\end{abstract}

\maketitle

\section{Introduction}
Although DNA is the chemical basis of the genetic code \cite{Crick1970-nm}, the  mechanical properties of DNA itself play a central mechanistic role in many cellular processes~\cite{Basu2021-ze,Finzi1995-ra, Gelles1998-ev,Wiggins2006-jr,Simpson_undated-vx}. DNA can be modeled as an elastic rod with twist and bend degrees of freedom subject to thermal fluctuations~\cite{Bates2005-ve}. 
On long length scales, the DNA conformation is described by the wormlike chain model (WLC) which depends on a single parameter, the persistence length $L_p\approx 50$\;nm~\cite{Bustamante2000-yx}.
However, the success of the WLC model on these long length scales does not imply that it is applicable at shorter, biologically-relevant length scales~\cite{Garcia2007-yu,Wiggins2006-jr}. In fact, both atomic-force-microscopy (AFM) imaging of DNA molecules adsorbed to mica~\cite{Wiggins2006-ox} and cyclization measurements of molecules shorter than $L_p$ appear to demonstrate that DNA is much more flexible on short scales than predicted by the WLC model~\cite{Cloutier2004-eo,Geggier2010-jr,Vafabakhsh2012-hu}. 

In spite of this previous experimental work, significant questions remain. 
The DNA cyclization approach suffers from several shortcomings: There remains debate in the literature about both (i) the measured $J$ factor values as well as (ii) about subtle and technical biochemical assumptions implicit to the interpretation~\cite{Du2005-wv}. Even if these cyclization results are sound, (iii) the mathematical dependence of the observable $J$ factor on the underlying DNA mechanics model is highly non-trivial. Although it is straightforward to determine whether a DNA mechanics model predicts the observed $J$ factor, the observed $J$ factor is insufficient to determine the DNA mechanics model~\cite{Drozdetski2019-cx}.  We therefore classify the cyclization approach as an \textit{indirect} measure of DNA mechanics. In contrast, we would classify AFM studies as a direct measurement of the DNA bending energy as a function of curvature; however, the interpretation of these experiments are subject to an assumption about whether the surface-adsorbed DNA molecules maintain the same mechanical properties as DNA in a physiologically relevant buffer~\cite{Wiggins2006-ox}. Can a direct measurement of the DNA torque-bend relation be measured in solution? Single-molecule techniques have been developed to directly measure the torque-twist relation; however, no equivalent methods have yet been developed to directly measure the torque-bend relation~\cite{Strick2000-er,Gore2006-gx,Neuman2010-oi}.

\begin{figure*}[ht!]
    \begin{center}
     \includegraphics[scale=0.7]{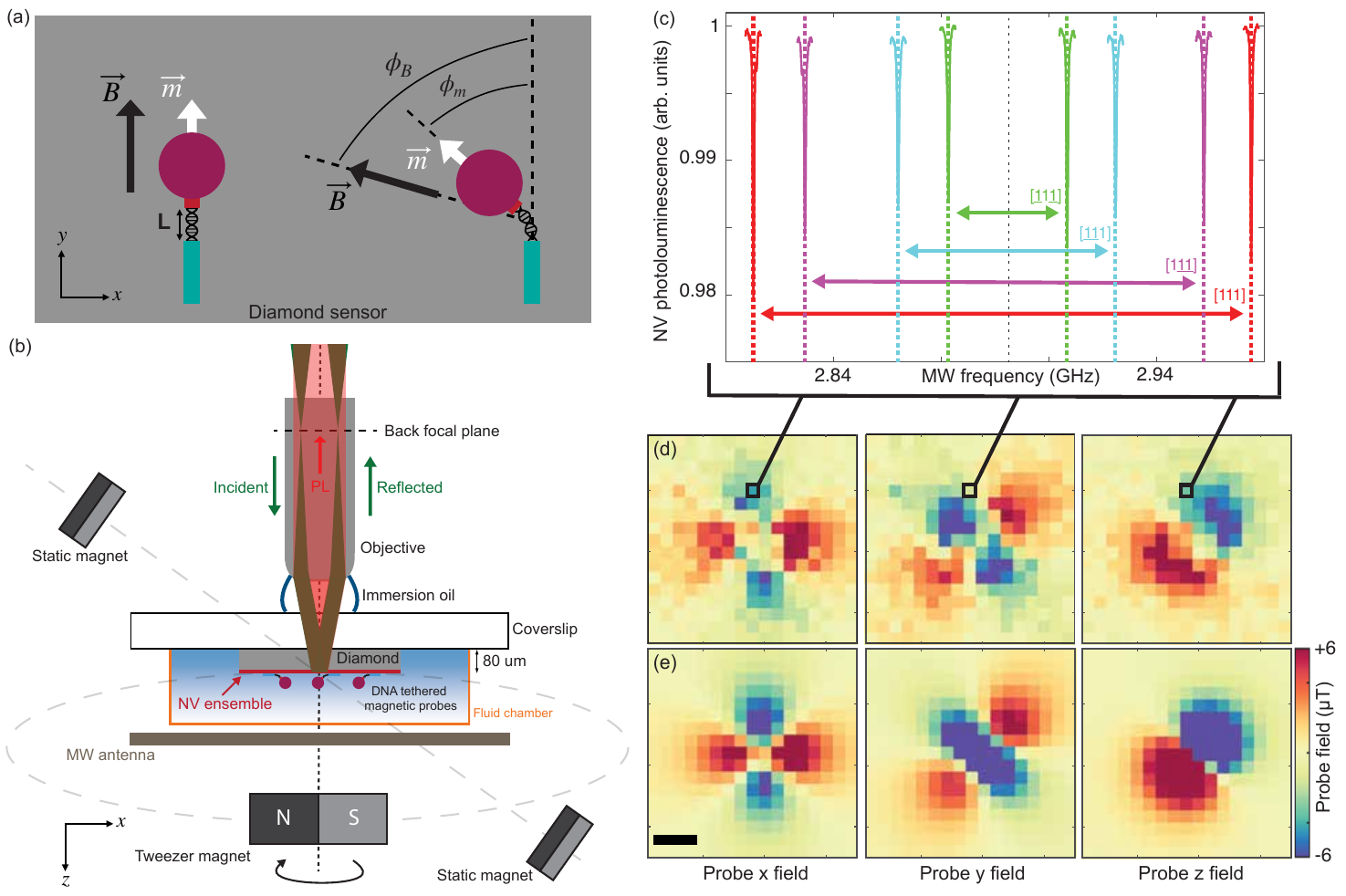}
     \end{center}
\caption{\textbf{Nano-mechanical torque-balance.} (a) Top-down schematic view of a single DNA molecule attached at one end to a diamond sensor and to a magnetic nanoparticle probe at the far end. An applied magnetic field $\vec{B}$ exerts a torque on the probe magnetic moment $\vec{m}$. The torque exerted by the applied magnetic field on the probe is balanced by the bending torque exerted by the DNA on the probe. (b) Schematic of the bio-compatible wide-field magPI platform. An 80\;\textmu m thick diamond sensor has near-surface NV defects in the top 150\;nm, and DNA tethered magnetic probes attached to the surface. The diamond is attached to a glass coverslip and embedded in a fluid chamber. Incident green laser light excites the NV layer in a total-internal-reflection geometry. The photoluminescence (PL) emitted by the NVs is collected and imaged onto the camera (not shown). Microwave (MW) excitation of the NVs is delivered by a broadband MW antenna from below. Two static magnets exert a magnetic field along the [111] NV orientation. A tweezer magnet placed along the optical axis is rotated in the $xy$ plane and exerts a magnetic field in the plane of the diamond. (c) Optically-detected-magnetic-resonance (ODMR) spectrum illustrating the four Zeeman splittings of the four NV orientations (different colors) about the zero-field-splitting (black vertical dashed line). (d) Measured probe field maps. ODMR spectra are measured at each pixel enabling the construction of magnetic probe field images as discussed in the main text. (e) Simulated probe field maps. Fitting to simulated probe magnetic field vector projection images enables measurement of the probe moment vector. Scale bar is 2 \textmu m.}
\label{fig:schematic}
\end{figure*}

A torque-balance assay measures the torque response as a function of bending angle. 
The challenge of realizing the DNA torque-balance assay lies in measuring the small torque exerted by individual DNA molecules and of probing the bend response at short length scales. In this work, we construct a nano-mechanical torque-balance to address both of these challenges. A ferromagnetic nanoparticle probe attached to one end of a single DNA molecule is used to bend the DNA molecule and to probe the bending response of DNA. At the far end from the ferromagnetic probe, the DNA molecule is constrained to a diamond quantum magnetic field sensor [Fig.~\ref{fig:schematic}(a)]. An external applied magnetic field exerts a torque on the probe which is balanced by the torque exerted by the DNA molecule. The probe magnetic moment $\vec{m}$ and the applied magnetic field $\vec{B}$ are measured simultaneously using a wide-field quantum magnetic probe imaging (magPI) platform comprising a near-surface ensemble of nitrogen-vacancy (NV) quantum defects in the diamond sensor~\cite{Kazi2021-dy}. By changing the orientation and strength of the applied magnetic field and measuring the orientation of the magnetic probe, we directly measure the DNA bend response as a function of bend angle. This quantum-enabled measurement scheme allows direct access to the bend response of individual DNA molecules, building on prior experiments using magnetic tweezers~\cite{Strick1998-ee, Gore2006-gx}. 

Using the torque-balance assay, this paper aims to answer the following questions: (i) Can the torque exerted by DNA bending be measured directly? (ii) What is the energy requirement to bend $>$50\;nm DNA molecules and (iii) sub-50\;nm DNA molecules? The realization of the torque-balance assay affirmatively answers question (i). Additionally, we address question (ii) by measuring the bend response of a DNA molecule with contour length $L>L_p$. The bend response of this molecule is found to be consistent with the WLC model prediction, which is expected at the long length-scale~\cite{Wiggins2006-ox}, demonstrating a technique that can be used towards a definitive answer to question (iii). 

This paper is organized as follows. Section~\ref{methods} introduces the nano-mechanical DNA torque-balance experimental assay, considers the required experimental parameters to probe the DNA bend response, and describes the DNA torque-balance construction. Section~\ref{widefieldimaging} details the wide-field quantum magPI platform used to measure the probe moment vector and applied external magnetic field vector. Section~\ref{controls} shows the ability of the assay to distinguish between three distinct DNA-probe configurations: an immobilized probe, a torsion-free DNA-tethered probe, and a probe in a torque-balance configuration. Section~\ref{bendstiffness} presents the direct measurement of bend response as a function of bend angle for an individual 200\;nm DNA molecule. Section~\ref{conclusion} provides an outlook towards using the torque-balance assay to build a quantitative understanding of the bending of short DNA molecules.

\section{Nano-mechanical torque-balance}\label{methods}
\subsection{Experimental principle}
The nano-mechanical torque-balance [Fig.~\ref{fig:schematic}(a)] leverages the principle of mechanical equilibrium to measure the bend torque exerted by an individual DNA molecule. For a given applied field, the torques acting on the DNA-tethered magnetic probe are balanced:
\begin{equation}
    \vec{\tau}_{\textrm{DNA}}= \vec{\tau}_{B}
    \label{equation:torquebalance}
\end{equation}
where $\vec{\tau}_{\textrm{DNA}}$ is the torque exerted by the DNA on the probe and 
\begin{equation}
    \vec{\tau}_{B}=\vec{m}\times\vec{B}
    \label{equation:magnetictorque}
\end{equation} 
is the torque exerted by the applied magnetic field $\vec{B}$ on the probe with magnetic moment $\vec{m}$. The probe must be strictly ferromagnetic so that $\vec{m}$ is independent of $\vec{B}$. Additionally, $\vec{B}$ must be spatially homogeneous such that the applied field only applies a torque and not a force on the magnetic probe. 

Using the torque-balance, this work aims to test the applicability of the WLC model which treats DNA as a elastic rod with bend energy that is quadratic in bend angle~\cite{Wiggins2006-jr}:
\begin{equation}
    E_{\textrm{WLC}}=\frac{1}{2}k_B T\frac{L_p}{L}\phi^{2}_{\textrm{DNA}}
    \label{equation:WLC_energy}
\end{equation}
where $L$ is the DNA contour length, $\phi_{\textrm{DNA}}$ is the in-plane DNA bend angle and $k_B T$ is the thermal energy. This model predicts that the torque exerted by DNA is linear in bend angle:
\begin{equation}
    \tau_{z}(\phi_{\textrm{DNA}})= \frac{\partial E_{\textrm{WLC}}}{\partial\phi_{\textrm{DNA}}}=k_BT\frac{L_p}{L}\phi_{\textrm{DNA}}
    \label{equation:WLC}
\end{equation}
where $\tau_{z}$ is the out-of-plane DNA bend torque. The torque-balance assay directly interrogates the DNA bend-torque relation $\tau_{z}(\phi_{\textrm{DNA}})$ by measuring the probe moment vector as a response to varying applied field directions and magnitudes. If the DNA bend response is governed by the WLC, Eqs.\;\ref{equation:torquebalance}\;\&\;\ref{equation:WLC} can be used to measure the WLC persistence length $L_p$ for a given DNA length $L$. If the DNA bending is not described by the WLC model, the assay can illuminate the underlying DNA bending physics by measuring the functional dependence of bend torque on bend angle.

\subsection{Ferromagnetic probe chosen to maximize sensitivity to DNA torque}
To measure a bend response of the DNA molecule in the torque-balance assay, we must apply torques within the dynamic range of the assay. The torque dynamic range for a given DNA contour length is constrained by starting with the WLC prediction. Using Eq.~\ref{equation:WLC}, we predict that the torque required to bend a $L$\;=\;200\;nm DNA molecule by $\phi_{\textrm{DNA}}$\;=\;1\;rad is $\tau_{\textrm{WLC}}\approx$\;1\;pN\;nm. As detailed in the next section, applied magnetic fields on the order of 1\;mT are used in the torque-balance assay, constraining the probe magnetic moment that can be used to measure the DNA molecule's bend response to be approximately $10^{-18}$\;A\;m$^{2}$. We satisfy this low magnetic moment requirement by using single-domain bio-compatible cobalt nanoparticles (Turbobeads GmbH): assuming a uniform bulk magnetization, we estimate the magnetic moment of the (approximately 30\;nm) single-domain nanoparticles to be $2\times10^{-18}$\;A\;m$^{2}$. This moment magnitude enables a 2\;pN\;nm applied torque in a 1\;mT magnetic field, on the same order as $\tau_{\textrm{WLC}}$ and sufficient to probe the bend response of individual DNA molecules.

\subsection{DNA-ferromagnetic probe torque-balance assay construction}
For the DNA to exert a torque on the magnetic probe, we use ``constrained" DNA molecules with multiple attachment points between both the sensor surface and the DNA and the DNA and magnetic probe. These constrained DNA molecules are engineered with two distinct lock-and-key binding pairs: digoxigenin/anti-digoxigenin for the surface-DNA attachments and biotin/streptavidin for the DNA-probe attachments~\cite{Kovari2018-tv}. To form the constrained DNA constructs, a free DNA fragment (of programmable length) is ligated together with an oligo containing multiple digoxigenin labels at one end, and an oligo containing multiple biotin labels at the far end. The free DNA fragment can be made arbitrarily short between the surface-DNA and DNA-probe linking oligos, in principle allowing for constructs with nanometer-scale free DNA lengths. 

The protocol for assembling the DNA constructs on diamond (based on Refs.~\cite{Kovari2018-tv,Yang2016-is}) is outlined here, and the full protocol is given in the Supplemental Material (SM)~\cite{Kazi2024-pp}. The diamond sensor is mounted to a glass coverslip and integrated into a fluid chamber with inlet and outlet ports. The construct reagents are introduced in solution one-by-one into the chamber and incubated for fixed time intervals. The un-bound excess is rinsed out using a phosphate-buffered-saline-based buffer. First, anti-digoxigenin is flowed in and non-specifically bound to the diamond surface. Then, a casein buffer is used to block the rest of the diamond surface to prevent non-specific binding of excess DNA molecules and magnetic probes. After surface blocking, the engineered DNA constructs are introduced and the digoxigenin ends bind to the anti-digoxigenin tethering sites at the diamond surface. Next, streptavidin-coated ferromagnetic probes are introduced that bind to the biotin end of the DNA molecules. Finally, the chamber is rinsed with experimental buffer. 


\begin{figure*}[ht!]
    \begin{center}
     \includegraphics[scale=0.85]{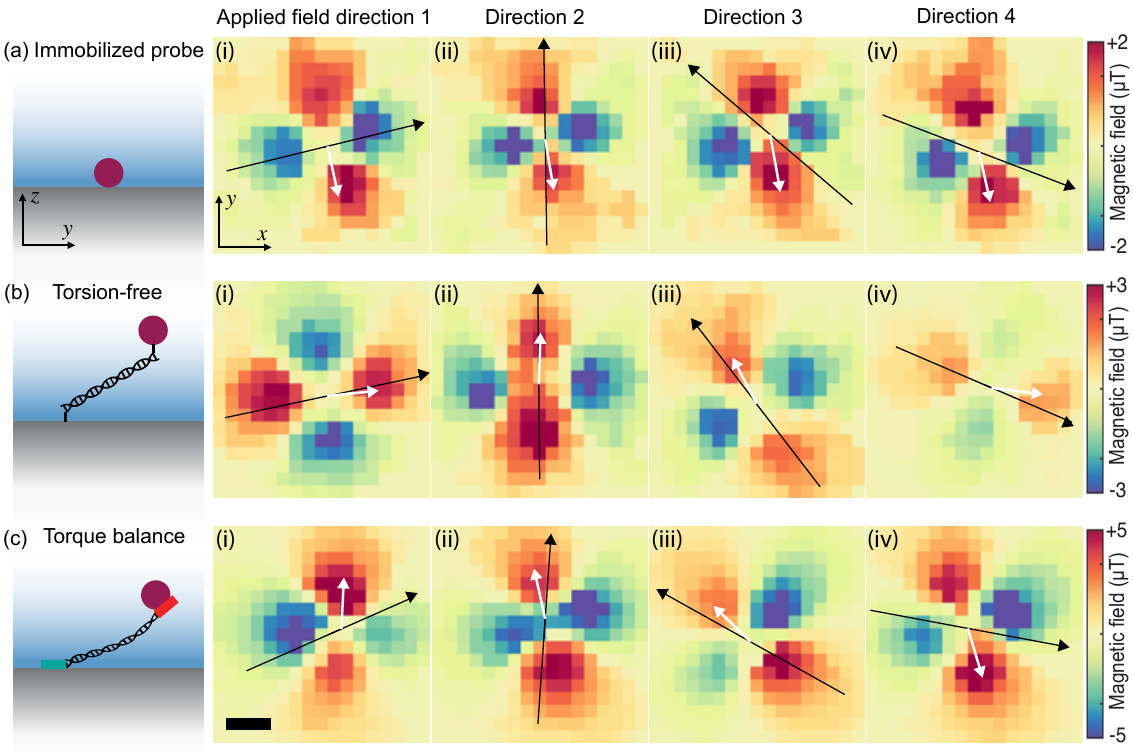}
     \end{center}
\caption{\textbf{Probe orientation illuminates qualitative differences in DNA binding character.} Ferromagnetic probe field images with changing applied field angle are shown for three different biomechanical configurations in the three rows, and four different applied field directions in the four columns. In each series, the dipole image is shown, overlaid with the fitted moment direction vector (white, shorter) and the applied field direction vector (black, longer). Scale bar is 2 \textmu m. (\textbf{a}) Probe immobilized on the diamond sensor surface. The dipole moment direction is independent of applied field angle and points in a constant direction. (\textbf{b}) Torsion-free DNA tethered probe. The probe moment points along the applied field direction. (\textbf{c}) Probe in a nano-mechanical torque-balance. The probe moment orientation is given by a balance between a torque from the applied field and a torque from the DNA. The probe aligns with the field for certain angles of the magnetic field and deviates from the applied field from other angles due to the torque applied by the DNA.}
\label{fig:comparison}
\end{figure*}

\section{magPI measures both probe and applied magnetic fields}\label{widefieldimaging}
An essential component of the torque-balance assay is measurement of the applied magnetic torque on the DNA-tethered probe, requiring a technique that measures both the probe magnetic moment vector and applied magnetic field vector. The simultaneous measurement of these two vectors is enabled by the magPI platform comprising a high-density ensemble of NV defects in the top 150\;nm of the diamond~\cite{Kleinsasser2016-jp,Kazi2021-dy}. A schematic of the bio-compatible platform is shown in Fig.~\ref{fig:schematic}(b). Magnetic probes are tethered by DNA molecules to a diamond sensor in a fluid chamber, and the NV defect electron spins are controlled with laser and microwave (MW) excitation. We image the emitted NV photoluminescence (PL) onto a sCMOS camera to enable wide-field vector magnetometry~\cite{Levine2019-xt}. Further details about the magPI platform can be found in Sec.~2 in the SM~\cite{Kazi2024-pp}.

Both probe and applied magnetic fields cause Zeeman splitting of the electron spin energy levels of the NV defects. Each camera pixel records the PL from several hundred NV centers that are oriented in four possible crystallographic directions~\cite{Abe2018-wm}. From each pixel we measure an optically-detected-magnetic-resonance (ODMR) spectrum by sweeping the MW frequency and measuring the optical response. An example ODMR spectrum shown in Fig.~\ref{fig:schematic}(c) illustrates the four Zeeman projections associated with each NV orientation. Measuring the ODMR spectrum in wide-field enables mapping of the four Zeeman projections across an imaging field of view~(Sec.~2.3 in SM~\cite{Kazi2024-pp})~\cite{Fescenko2019-vd}. These Zeeman maps are linearly transformed to maps of the lab-frame magnetic field vector components [Fig.~\ref{fig:schematic}(d)]. 

To perform this transformation, each Zeeman splitting must be assigned to a specific NV orientation, which in an arbitrary field is challenging due to the $C_{3v}$ symmetry of the NV defect~\cite{Abe2018-wm}. We break this symmetry by using an applied field comprising two superposing magnetic fields: a ``static" field parallel to the [111] NV orientation and a rotating ``tweezer" field parallel to the diamond plane [Fig.~\ref{fig:schematic}(b)]. This asymmetric total applied field with millitesla-magnitude enables each of the four Zeeman projections to be uniquely assigned (Sec.~2.4 in SM~\cite{Kazi2024-pp}). The total applied field is not fully in-plane as suggested by Fig.~\ref{fig:schematic}(a) because the static field has a small out-of-plane component.

The NVs measure both probe and applied magnetic fields, but these fields can be disentangled since the millitesla-scale applied field is homogeneous over the imaging field of view. The applied field is obtained by averaging the measured magnetic field across the image, and the microtesla-scale probe dipole field maps are obtained by subtracting off the applied field at each pixel. 

To speed up data acquisition, which as will be seen is critical to minimize phototoxicity of the DNA tethers, we further exploit the mismatch in magnitudes between applied field and probe fields. The Zeeman splittings caused by the applied field are on the order of tens of MHz, while the Zeeman splittings caused by probe fields are tens to hundreds of kHz. We thus first measure the ODMR frequencies for a given applied field direction and magnitude in the absence of DNA-tethered magnetic probes [Fig.~S5 in SM~\cite{Kazi2024-pp}], then during measurement of DNA tethers we only apply MW frequencies in a 3~MHz window around each ODMR for each applied field configuration [Fig.~S6 in SM~\cite{Kazi2024-pp}]. 

The measured probe field lab-frame components [Fig.~\ref{fig:schematic}(d)] are fit by a least-squares method to a six-parameter magnetic dipole model [Fig.~\ref{fig:schematic}(e)] that enables determination of the probe position and magnetic moment vectors (Sec.~3 in SM~\cite{Kazi2024-pp}). As illustrated in the probe fields of Figs.~\ref{fig:schematic}(d)\;\&(\;e), there is a discrepancy between the center regions of the experimental and modelled probe field images. This discrepancy exists because the probe dipole produces high magnetic field gradients that are unable to be imaged by the magPI platform. The large magnetic field gradients inhomogenously broaden the ODMR spectra, reducing ODMR contrast and washing out the probe field signal. To fit the probe moment vector, these high-gradient regions are masked (Sec.~3.1 in SM~\cite{Kazi2024-pp}). Errors associated with probe field fitting are discussed in Sec.~3.2 in the SM~\cite{Kazi2024-pp}.

\section{Differentiation between torsion-free and constrained DNA molecules}\label{controls}
We first characterize the precision of the torque-balance assay by measuring the probe response of a series of constructs with known mechanical properties. These experimental tests validate the torque-balance assay's ability to measure probe orientation, and to discern distinct bio-mechanical configurations by their responses to the applied field. The three experiments presented here examine the probe response to the applied field in the case of an immobilized probe, a torsion-free DNA-tethered-probe, and a constrained DNA-tethered probe like those used in the torque-balance assay. 

In the first experiment, the magnetic probe is immobilized on the diamond sensor surface [Fig.~\ref{fig:comparison}(a), left] in the absence of DNA. The probe magnetic moment is fixed by the surface interaction, independent of the applied field. This experiment provides a non-trivial experimental test: if the probe moment is found to reorient with applied field, then the probe may not be purely ferromagnetic and the platform is unable to measure both probe and applied fields. The immobilized probe response to applied field is shown in the image panels in Fig.~\ref{fig:comparison}(a), where the four images correspond to four different applied field directions. The in-plane probe magnetic moment vector is overlaid in white and the in-plane applied field vector is overlaid in black. As expected, for the immobilized probe, the measured probe moment vector points in an arbitrary fixed direction, independent of applied field. This result validates the magPI platform's ability to sense both probe and applied fields and demonstrates probe ferromagnetism, crucial in order for probe direction to be correlated with DNA orientation.

\begin{figure*}[ht!]
    \begin{center}
     \includegraphics[scale=1]{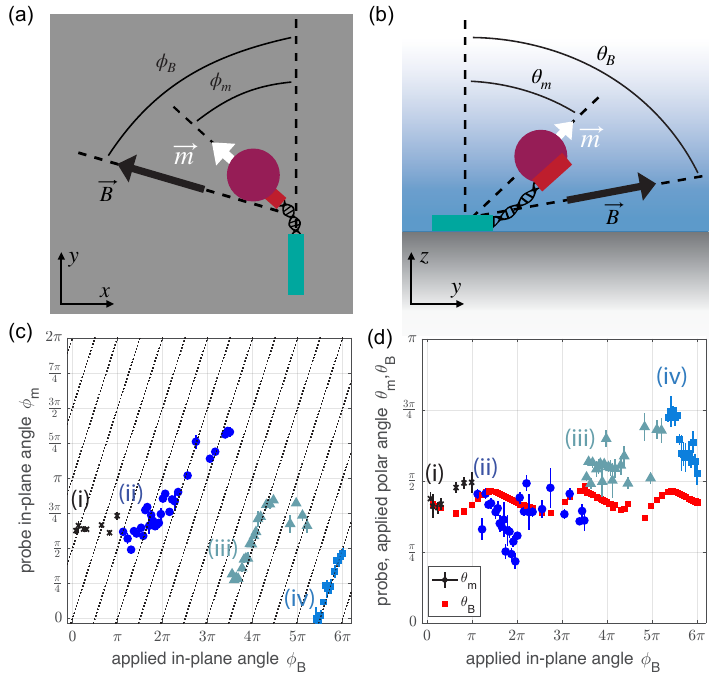}
     \end{center}
\caption{\textbf{Deflections in probe orientation caused by DNA bending torque.} (a) Top-down schematic of torque-balance assay showing with moment vector in white, applied field vector in black, and moment and applied field in-plane angles ($\phi_\textrm{m}$,$\phi_\textrm{B}$) labeled. (b) Side-view schematic of torque-balance assay showing with moment vector in white, applied field vector in black, and moment and applied field polar angles  ($\theta_\textrm{m}$,$\theta_\textrm{B}$) labeled. (c) Probe moment in-plane angle $\phi_\textrm{m}$ as a function of applied field in-plane angle $\phi_\textrm{B}$ for a 200\;nm DNA tether. Dashed lines with unit slope are plotted and represent the response of a torsion-free DNA tether. Deviations from dashed lines indicate the DNA is exerting a torque on the magnetic probe. Four trajectories (i)-(iv) are observed in which the DNA tether undergoes a distinct response. (d) Probe moment polar angle $\theta_\textrm{m}$  and applied field polar angle $\theta_\textrm{B}$ as a function of applied field in-plane angle $\phi_\textrm{B}$. For a torsion-free tether, $\theta_\textrm{m}\approx\theta_\textrm{B}$, deviations between the probe direction and the applied field direction indicate the DNA is exerting a torque on the magnetic probe.}
\label{fig:600bpangles}
\end{figure*}

In the second experiment, the probe is tethered by a torsion-free DNA molecule with a single digoxigenin and biotin marker at each end, meaning the surface-DNA and DNA-probe are attached by a single binding site [Fig.~\ref{fig:comparison}(b), left]~\cite{Kovari2018-tv,Kazi2021-dy}. The torsion-free DNA tethered magnetic probe is thus free to align to the applied field. This experiment provides an additional set of experimental tests: if the torsion-free tethered probe is found to have a similar response to the immobilized probe, the DNA tethers are unable to be formed on diamond. Additionally, if the probe moment is found to deviate from the applied field direction, the DNA may be cooperatively adhering to the diamond surface. The torsion-free tethered probe response to applied field shown in Fig.~\ref{fig:comparison}(b) shows that the probe moment aligns to the applied field. This experiment demonstrates the ability of the magPI platform to effectively measure field-dependent probe orientation and validates the diamond-DNA tethering protocol.

In the third experiment, the torque-balance assay, the probe is tethered by a constrained DNA molecule [Fig.~\ref{fig:comparison}(c), left]. This experiment tests whether the assay can detect the multiple binding sites used with constrained tethers: if the response of the probe in this experiment is different from both the first two experiments, the assay is able to detect DNA back-action on the magnetic probe. The four images shown in Fig.~\ref{fig:comparison}(c) show a different magnetic probe response to the changing applied field direction compared to Fig.~\ref{fig:comparison}(a) and Fig.~\ref{fig:comparison}(b). In Fig.~\ref{fig:comparison}(c)(i), the probe is oriented away from the applied field direction, similar to the immobilized probe. Then, in Fig.~\ref{fig:comparison}(c)(ii)\&(iii), the probe is oriented along the applied field direction, similar to the torsion-free probe. Finally, in Fig.~\ref{fig:comparison}(c)(iv) the probe is again oriented away from the applied field. The probe moment direction in this experiment is determined by both the torque exerted by the external field and the torque exerted by the DNA on the magnetic probe. As will be seen in the next section, the DNA bend torque magnitude can be quantified by measuring the deflection between probe orientation and applied field direction. 


These three experiments demonstrate the assay's ability to determine probe orientation, validate ferromagnetism of the probes, and observe back-action of the DNA on the magnetic probe due to bending torque.

\section{Direct measurement of the torque generated by DNA bending}\label{bendstiffness}
We obtain the bend response of a single 200\;nm DNA molecule by measuring the probe response to a series of closely-spaced applied field angles. For each applied field angle, we measure the probe moment vector. The magnetic torque on the probe depends on the applied field vector and probe moment vector, as shown in Eq.~\ref{equation:magnetictorque}. We thus first characterize the probe orientation defined by probe in-plane angle $\phi_\textrm{m}$ [Fig.~\ref{fig:600bpangles}(a)] and probe polar angle $\theta_\textrm{m}$ [Fig.~\ref{fig:600bpangles}(b)] over three full in-plane revolutions of the applied field. Fig.~\ref{fig:600bpangles}(c) shows $\phi_\textrm{m}$ as a function of applied field in-plane angle $\phi_\textrm{B}$. Several distinct probe trajectories are observed. In the first trajectory [Fig.~\ref{fig:600bpangles}(c)(i)], the probe in-plane angle is relatively unchanging such that the probe appears immobilized, potentially caused by stochastic binding of the tether to the surface. In the second trajectory [Fig.~\ref{fig:600bpangles}(c)(ii)], after about half a revolution, the probe begins to follow the field, however, the slope of the trajectory is less steep than the torsion-free tether response (dashed lines). This indicates that the DNA is bending and the both the DNA and applied field are exerting a torque on the probe. After about a full revolution of applied field [Fig.~\ref{fig:600bpangles}(c)(iii)], the probe discontinuously jumps in orientation before resuming a near-linear response with applied field. After a final jump [Fig.~\ref{fig:600bpangles}(c)(iv)], the probe follows the dashed lines, appearing completely torsion-free, after which the probe signal vanishes. This suggests that the surface-DNA binding changes between (iii) and (iv) from a torque-balance to a torsion-free configuration, then laser-induced phototoxicity un-tethers the DNA or probe. The discrete jumps in $\phi_\textrm{m}$ are given by the probe reorienting in both the in-plane and polar direction $\theta_\textrm{m}$ as seen in Fig.~\ref{fig:600bpangles}(d). 

\begin{figure*}
    \begin{center}
     \includegraphics[scale=1]{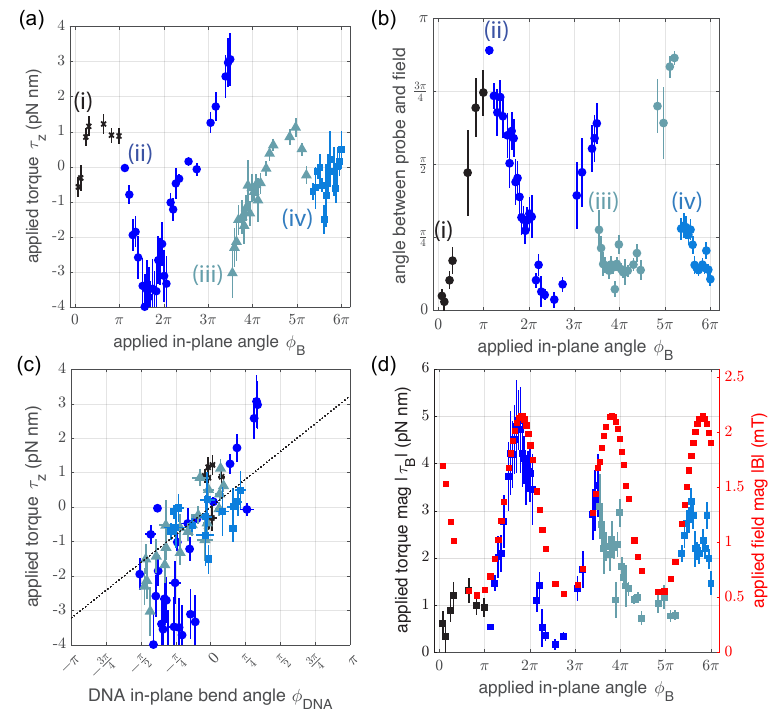}
     \end{center}
    \caption{\textbf{Response of dipole to applied field measures bend response of 200\;nm DNA molecule.} (a) Out-of-plane applied torque $\tau_z$ as a function of applied field in-plane angle $\phi_\textrm{B}$. Discontinuities in out-of-plane applied torque represent when the DNA buckles under a high applied torque as seen in the discontinuities in angle between probe and field. (b) Angle between probe moment vector and applied field vectors as a function of applied field in-plane angle $\phi_\textrm{B}$. This angle scales with the magnitude of applied torque on the magnetic probe. (c) Out-of-plane magnetic torque $\tau_{z}$ as a function of DNA in-plane bend angle $\phi_{\textrm{DNA}}$, with the colors from each trajectory maintained. Dashed line is the WLC prediction with $L_p$\;=\;50\;nm and $L$\;=\;200\;nm. The data shows reasonable agreement with the WLC model. (d) Magnitude of applied torque $|\tau_B|$ and applied field magnitude $|B|$ as a function of applied field in-plane angle $\phi_\textrm{B}$. The applied field magnitude oscillates as a function of applied field angle because of the two superposing magnetic fields as discussed in Section~\ref{widefieldimaging}. The applied torque magnitude scales with applied field magnitude.}
    \label{fig:600bpbendstiffness}
    \end{figure*}

Using probe in-plane angle $\phi_{\textrm{m}}$, polar angle $\theta_{\textrm{m}}$ and the probe moment magnitude, we calculate the applied torque on the probe. For this DNA tether, the fitted probe magnitude is $2.3\times10^{-18}$\;A\;m$^2$, consistent with the predicted moment magnitude necessary to probe the DNA bend response (Sec.~\ref{methods}B). The out-of-plane component of applied torque $\tau_z$ is plotted as a function of applied field in-plane angle in Fig.~\ref{fig:600bpbendstiffness}(a). In the first trajectory [Fig.~\ref{fig:600bpbendstiffness}(a)(i)], $\tau_z$ is low because the angle between probe vector and applied field vector is small. In the second trajectory, [Fig.~\ref{fig:600bpbendstiffness}(a)(ii)], $\tau_z$ shows an approximately linear relationship with applied in-plane angle, indicating the influence of DNA bending. Between the second and third trajectories [Fig.~\ref{fig:600bpbendstiffness}(a)(ii) and (iii)], $\tau_z$ flips sign but maintains its magnitude of approximately 3\;pN\;nm. Then, $\tau_z$ linearly scales for the remainder of the third trajectory, again representing the response of DNA bending. In the fourth trajectory [Fig.~\ref{fig:600bpbendstiffness}(a)(iv)], $\tau_z$ is measured to be approximately zero for the remainder of the trajectory, consistent with the torsion-free response seen in Fig.~\ref{fig:600bpangles}(c)(iv). 

Using Eq.~\ref{equation:WLC}, we can quantify the DNA bend torque response by measuring the out-of-plane torque as a function of DNA in-plane bend angle,  i.e. $\tau_z(\phi_{\textrm{DNA}})$. $\phi_{\textrm{DNA}}$ for each trajectory is measured by subtracting the probe in-plane angle from a ``home" orientation, so $\phi_{\textrm{DNA}}\;=\;\phi_{\textrm{m}}-\langle\phi_j\rangle$. For each trajectory (i)-(iv), the ``home" orientation $\langle\phi_j\rangle$ is taken as the probe angle for which $|\tau_z|$ is minimized. The applied torque $\tau_z$ is plotted as a function of DNA bend angle $\phi_{\textrm{DNA}}$ in Fig.~\ref{fig:600bpbendstiffness}(c). Additionally, the WLC model prediction (Eq.~\ref{equation:WLC}) with $L$\;=\;200\;nm and $L_p$\;=\;50\;nm is plotted as a dashed line. 

For a 200\;nm DNA tether, because $L>L_p$, the hypothesis is the bending should be described by the WLC model prediction~\cite{Wiggins2006-jr}. The data in Fig.~\ref{fig:600bpbendstiffness}(c) shows reasonable alignment with the WLC model prediction for $-\pi/4\leq\phi_{\textrm{DNA}}\leq\pi/4$, validating the torque-balance assay's ability to measure DNA bend response. Interestingly, we do not observe in-plane bend angles above $\pi/2$. This is due to the polar-angle reorientation of the probe moment vector at large applied torque, like between trajectories (ii) and (iii). However, we can observe in trajectory (ii) and to a lesser extent in trajectory (iii) that even before the reorientation there is a deviation between the WLC prediction and the bend response. This deviation could suggest an energetic model of DNA bending that is distinct from the WLC model, for example, a bend energy model that is quadratic for low bending angles but ``turns on" a linear energy term for high bending angles~\cite{Wiggins2006-jr,Drozdetski2019-cx}. For a model of this type, the DNA bend-torque relation would be linear around zero bend angle and saturate for some critical bend angle. In order to fully understand the DNA bending, further measurements with varying applied torque magnitudes are required as discussed in the next section. In all experiments, the magnetic probe becomes un-tethered after minutes to hours under optical illumination.

As a consistency check, we show the total applied torque magnitude on the probe $|\tau_B|$ overlaid with the the applied field magnitude $|B|$ in Fig.~\ref{fig:600bpbendstiffness}(d), which qualitatively scale together. However, the four distinct trajectories show differing total applied torques, corroborating the analysis that the DNA configuration changes between each trajectory. Additionally, the angle between the probe vector and applied field vector shown in Fig.~\ref{fig:600bpbendstiffness}(b) shows a changing probe response to applied torque in each trajectory. 

This direct measurement of the bend response of a 200\;nm DNA molecule is validated by the low-bend-angle agreement with the WLC prediction. Additional higher-stiffness and torsion-free DNA tether torque responses are provided in Sec.4 of the SM~\cite{Kazi2024-pp}. These additional measurements illustrate the assay's ability to characterize varying DNA bend responses, and also highlight the challenge of interpreting this response at the present state of this technology.

\section{Conclusion and Outlook}\label{conclusion}
In this work, a nano-mechanical torque-balance assay is used to measure the bend response of individual DNA molecules. A wide-field quantum magnetic probe imaging platform simultaneously measures both the vector magnetic field produced by DNA-tethered ferromagnetic probes and the applied magnetic field vector. The probe dipole magnetic field images are fit to measure the probe moment vector, and the platform is able to distinguish three distinct biomechanical configurations: immobilized probe, torsion-free DNA tether and constrained, multiply-bound DNA tether. For a 200\;nm DNA molecule, we directly measure the bend response and find it to be consistent with the WLC model with $L_p$\;=\;50\;nm. 

This proof-of-principle demonstration of the torque-balance assay opens up the possibility of quantitative measurements of short DNA bending. The first measurement would be a controlled DNA molecule length-dependence study to identify the length scales over which the WLC model holds. The torque-balance DNA constructs used can be made arbitrarily short between the surface-DNA and DNA-probe linking oligos, allowing for measurements down to the nanometer length scale. Additionally, the assay could be used to measure sequence-dependent bending stiffness of individual DNA molecules~\cite{Geggier2010-jr} to provide insights about the effect of DNA form on function. However, to realize these controlled studies, an orthogonal imaging modality to validate the DNA constructs should be implemented. The control experiments in this work are able to differentiate between a torsion-free and torque-balance DNA tether, but with a separate structure imaging modality such as the dark-field tethered-particle-motion (DF-TPM) assay~\cite{Milstein2011-kr} the length of each DNA construct may be able to be measured independently. Using the DF-TPM assay to quantify DNA tether length is conceptually simple but is challenging to realize, as it requires a highly sensitive imaging apparatus to measure the low amount of probe-scattered-light at high enough speeds to capture the sub-millisecond DNA tether correlation time dynamics with sufficient signal-to-noise.

Several areas for further sensor development should be addressed to realize large-scale quantitative DNA bending measurements. 
While the use of magnetic probes circumvents bleaching of fluorescent probes, a significant challenge in this experiment is excitation-laser induced phototoxicity of DNA tethers which limits the amount of time (and number of applied magnetic field configurations) under which the tethers can be studied, even with total-internal-reflection excitation. In this work, we found the multiply-bound tethers in the torque-balance configuration to be more robust than the singly-bound torsion-free tethers, but both kinds of DNA tethers were eventually subject to phototoxicity, while the immobilized probes were robust to high-intensity and long time-scale laser illumination. Solutions for using sufficient laser power to excite the NV defects while limiting tether destruction are key to enable long-time-scale, high-sensitivity measurements.

Improving the yield of bound DNA tethers (approximately 1-3 per (50\;\textmu m)$^2$ in this work) is also necessary, and may require new diamond surface preparation techniques such as with atomic layer deposition (ALD) of a different material~\cite{Xie2022-ay} or direct surface-functionalization~\cite{Rodgers2023-ny}. We found the DNA tether yield to be significantly higher on glass relative to diamond. To increase yield on diamond, ALD of alumina was done on one of two identical diamonds used in this work to try to mimic a glass surface. The DNA tether yield on the diamond surface after alumina ALD was not found to be statistically significant, so further explorations of surface preparation to increase DNA tether yield are warranted.

Next, we note that due to the single-pixel magnetic field gradients near the probe (Sec.~\ref{methods}), we are unable to utilize a majority of probe signal to determine probe orientation and distance. A promising approach to reduce the pixel size and thus mitigate single-pixel inhomogeneous broadening is to use super-resolution magnetic field imaging~\cite{Mosavian2023-sc}. This technique would increase sensitivity and spatial resolution of the probe field images, allowing for faster imaging with lower error on probe orientation fitting.

Another challenge is related to the vector magnetometry used in this work, which requires assigning each of the four measured Zeeman splittings to their corresponding NV crystallographic orientations. In the presence of small and symmetric magnetic fields, the orientation assignment is difficult because the ODMR spectra for the different NV orientations overlap. In this work, vector magnetometry is accomplished by applying a millitesla-scale asymmetric magnetic field. By using Fourier optical processing~\cite{Backlund2017-cr}, the individual NV orientations can in principle be imaged separately using downstream optical components, which would provide several advantages. First, the Zeeman-to-NV-orientation assignment would always be known, unambiguously determining lab-frame magnetic field without calibration. Second, Fourier processing would omit the need for an asymmetric applied field. This would allow a single magnet to supply the applied field, making the applied field magnitude constant for differing field directions and significantly simplifying the interpretation of the DNA bend-torque response. Additionally, microtesla-scale applied fields could be used, enabling the use of larger moment probe magnets to apply the same magnitude torques on the DNA molecules for faster and more accurate probe orientation imaging. 

In conclusion, we have developed a new biophysical assay that can directly measure the bend response of individual DNA molecules. This work combines the tethered-particle-motion assay, a long-standing workhorse of DNA single-molecule biophysics, with a magnetic tweezer and the high-sensitivity magnetic imaging enabled by NV quantum defects in diamond. With this proof-of-principle demonstration, we build towards a large-scale, quantitative measurement platform that could illuminate underlying DNA mechanics at biophysically-relevant short length-scales. More broadly, we show that advancements in quantum sensing technology may enable access to measurements of yet-unstudied fundamental biophysical phenomena.

\section*{Acknowledgements}
We acknowledge Yuri Choe and David Bergsman for alumina atomic layer deposition, Kento Sasaki for supplying the microwave antenna used in this work, and Isaiah Kim for initial magnetic tweezer field calibration. This material is based on work supported by the National Science Foundation under Grant No.~1607869 and the University of Washington Materials Research Science and Engineering Center (MRSEC;~DMR-1719797). 

\bibliography{main}

\end{document}


\pagebreak
\begin{center}
\textbf{\large Supplemental material: Direct measure of DNA bending by quantum magnetic imaging of a nano-mechanical torque balance}
\\
Z. Kazi et al.
\end{center}

    \section{Torque-balance assay construction protocol}
    The full protocol for assembling the torque-balance DNA tethers on diamond is detailed here. First, the diamond flow chamber is prepared, then the tethers are formed on diamond.
    \subsection{Diamond flow chamber}
    The diamond sensor (approx 80~\textmu m thick) is first cleaned by soaking in acetone and scrubbing with a cotton-tipped applicator. Then the diamond is sonicated in 1\% v/v Hellmanex III in milQ water for 30 mins, followed by sonication in milQ water for 30 mins. Next, the edges of the diamond are taped using Scotch double-sided tape (with the NV layer up) on a glass coverslip (Corning No. 1~1/2, after coverslips are sonicated for 30 min each in alconox, ethanol, and milQ water). After, a perfusion chamber (Grace Bio Labs CoverWell 622103) is placed over the diamond to create a sealed fluid chamber with inlet and outlet ports. 
    \subsection{DNA-tethered magnetic probe construction}
    The DNA tethers are assembled using a protocol adapted from Ref.~\cite{Kovari2018-tv}:
        \begin{itemize}
        \item Into prepared chamber, flow 100~\textmu L 1X phosphate buffered saline (PBS) (MilliporeSigma 806552-1L).
        \item Dilute anti-digoxigenin (Roche 11333089001) in PBS to 100~\textmu g/mL 
        \item Centrifuge anti-dig solution for 5 mins at 10,000 rpm.
        \item Flow anti-dig into chamber. Pipette to uniformly fill chamber. Use folded paper KimWipe at other side to draw fluid and exchange liquids in chamber.
        \item Sonicate 30 pM Turbobead solution, wait 60 mins.
        \item Flush chamber with 200~\textmu L PBS.
        \item Flush chamber with casein buffer (100~\textmu L) (WestEz Casein buffer), wait 30 mins.
        \item Flush chamber with 200~\textmu L DNA buffer ($\lambda$-buffer from~\cite{Kovari2018-tv}).
        \item Flow 20 pM DNA solution into chamber, wait 15 mins.
        \item Flush chamber with 200~\textmu L DNA buffer.
        \item Flow magnetic probe solution (10~\textmu L, using P20 with 10~\textmu L pipette tip). 
        \item Wait 15 mins. Longer wait times here results in clumping of the magnetic probes.
        \item Flush chamber with DNA buffer. (800~\textmu L, 4 x 200~\textmu L)
        \item Apply caps to flow chamber.
        \end{itemize}    
        The DNA tethers are robust over the flow chamber lifetime which is typically limited by fluid evaporation, causing bubbles to form in the chamber. Chambers are stored in a refrigerator at approximately 8$\degree$C between measurements.

    \section{Magnetic particle imaging (magPI) platform}
    \subsection{Diamond sensor details}
    The diamond sensor used in this work~\cite{Kazi2021-dy} comprises a near-surface, high density NV ensemble. A \SI{150}{\nano\meter} $^{15}$N doped, isotope-purified (99.999$\%$ $^{12}$C) layer was grown by chemical vapor deposition on an 80~\textmu m thinned electronic-grade diamond substrate (Element Six). After growth, the sample was implanted with 25~keV He$^{+}$ at $5 \times 10^{11}$~ions/cm$^{2}$ to form vacancies, followed by a vacuum anneal at 900~$\degree$C for \SI{2}{\hour} for NV formation and an anneal in $\text{O}_{2}$ at 425~$\degree$C for \SI{2}{\hour} for oxygen surface termination~\cite{Kleinsasser2016-jp}. The resulting ensemble has NV density approximately 0.1~ppm and ensemble spin coherence time $T^{*}_{2}$=2.5~\textmu s. Two such samples were prepared and used to take the data shown in the main text and Supplemental Material.

    \renewcommand{\thefigure}{S\arabic{figure}}
    \begin{figure*}[h]
        \centering
        \includegraphics[scale=0.35]{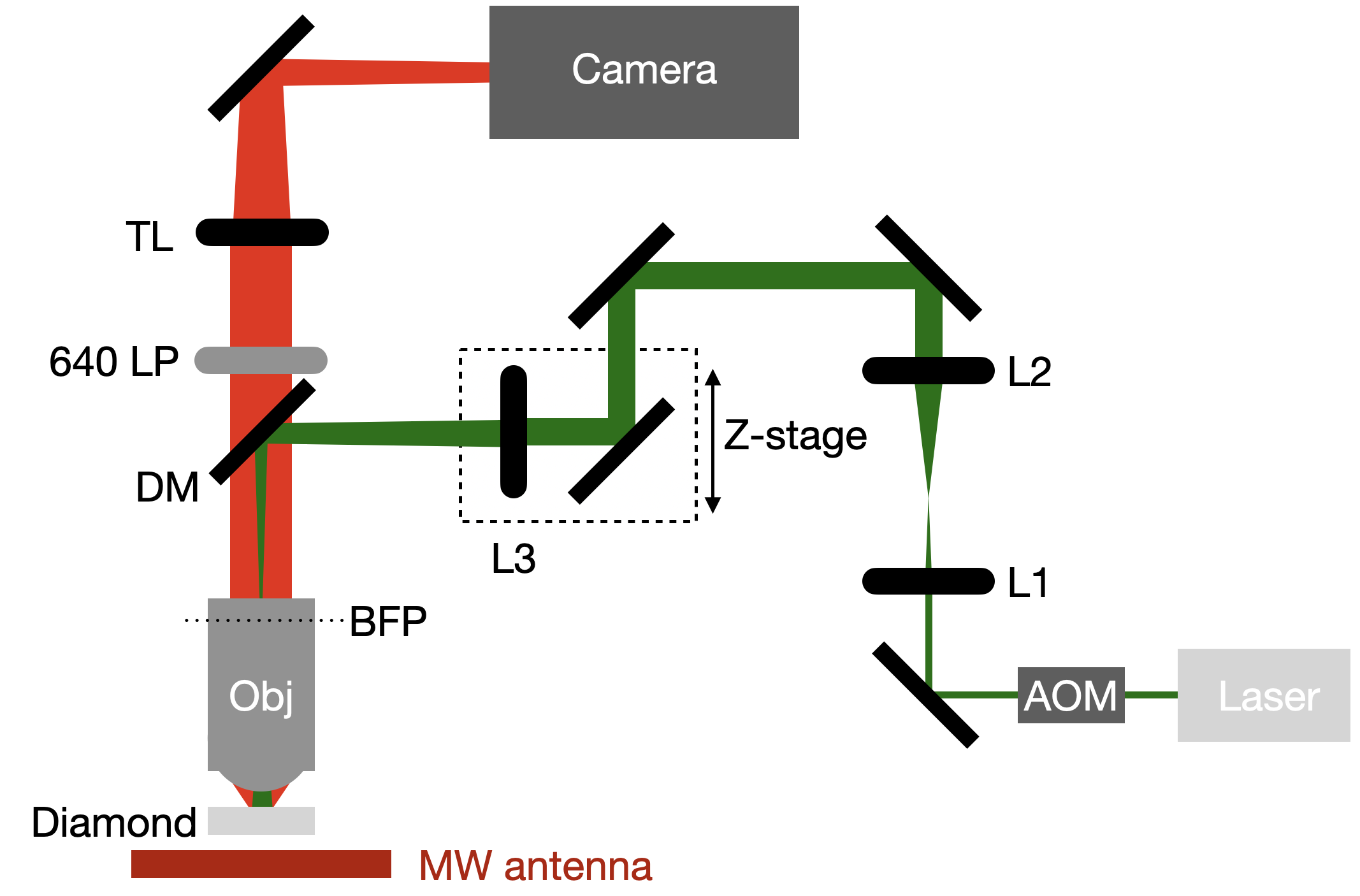}
        \caption{\textbf{DNA torque balance magnetic imaging experimental schematic.}}
        \label{fig:opticalschematic}
    \end{figure*}
    \subsection{Experimental set up}
    Fig~\ref{fig:opticalschematic} shows the DNA torque balance experimental schematic. Green (532~nm) light from a laser (LaserQuantum opus532 2W) passes through an acousto-optic modulator (AOM, Gooch \& Housego 3080-294). The laser beam is expanded by a telescope comprising lenses L1 and L2 and reflects off a mirror and passes through lens L3. L3 focuses the beam in the back focal plane (BFP) of the objective (2 mm from the back aperture of the Nikon objective). L3 and the closest mirror are placed on a translation Z-stage in order to translate the beam off axis for total-internal-reflection of the excitation beam at the diamond-fluid interface (see Fig.~1(b) in the main text). The excitation light reflects off a dichroic mirror (DM, Semrock LPD02-532RU-25) and excites the NVs with a collimated beam for wide-field imaging. The emitted fluorescence is collected by the objective lens (Obj, Nikon 60x 1.4 NA) and passes through the DM. The light is filtered using a 640 nm long pass (LP) filter, and focused by a tube lens (TL) onto a Hamamatsu OrcaFlash sCMOS camera.
    
    Microwave (MW) excitation drives transitions between NV electron spin states. The MW setup used in this work is designed to apply multi-toned MW pulses simultaneously to the NV ensemble to drive mutliple electron spin transitions and hyperfine transitions simultaneously. This is accomplished by mixing the MW signal from a WindFreak SynthNV that forms the LO of a mixer (Pasternack SFM2018) with a function generator supplying the IF to split the center MW tone into two tones split by 3.05~MHz that can drive both $^{15}$NV hyperfine resonances simultaneously and produce one combined resonance for each electron spin transition. The MW signal is then amplified (Minicircuits ZHL-16W-43-S+), passed through a circulator (Fairview SFC2040A) then sent to a broadband MW antenna~\cite{Sasaki2016-iv} with a resonance near the NV zero-field-splitting which excites the NV spins with a z-polarized MW magnetic field. The full ODMR spectrum with hyperfine-mixed driving is shown in Fig.~\ref{fig:fullODMR}.

    \renewcommand{\thefigure}{S\arabic{figure}}
    \begin{figure*}[h]
        \centering
        \includegraphics[scale=0.3]{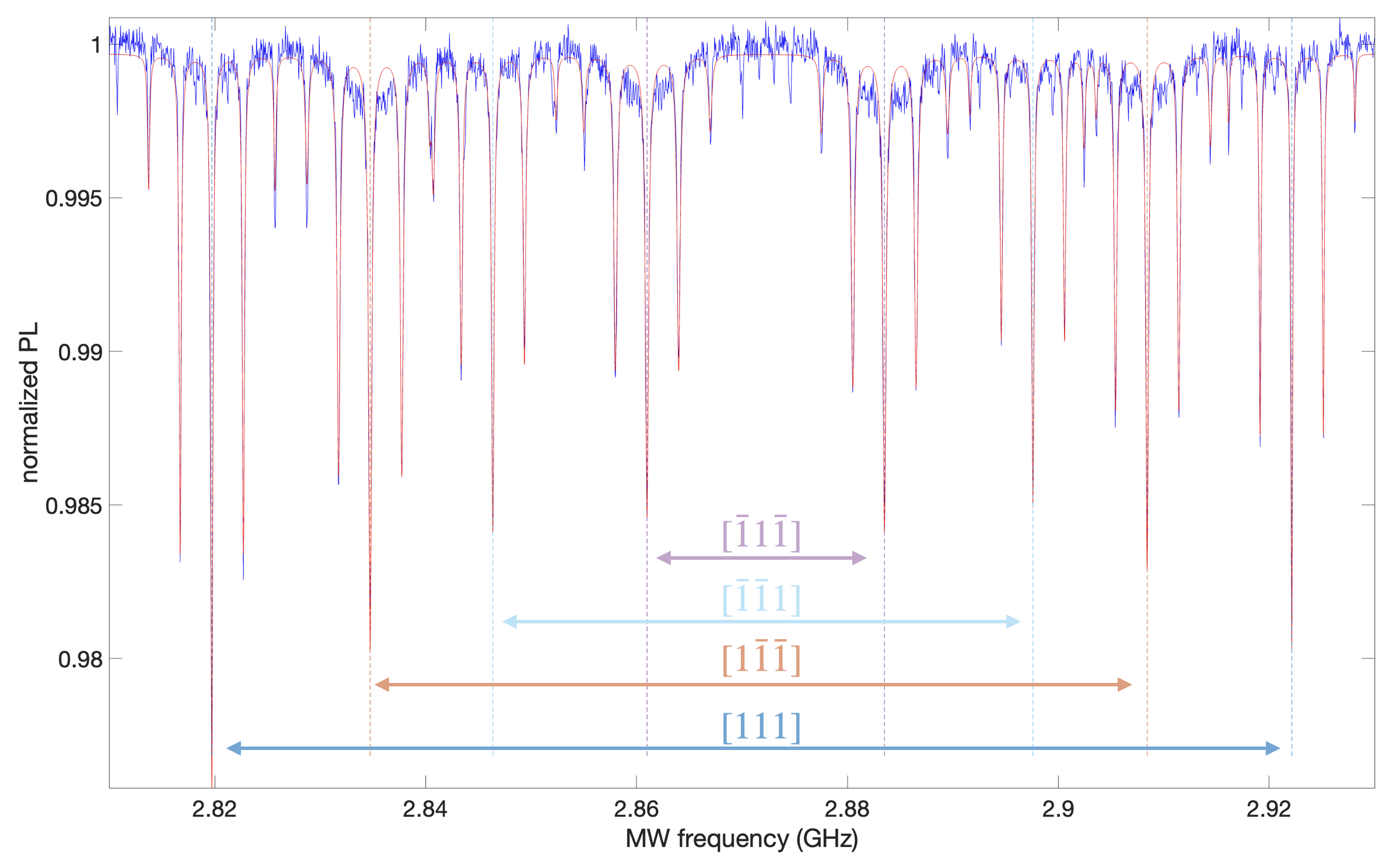}
        \caption{\textbf{Full ODMR spectrum with hyperfine-mixed MW driving.} The Zeeman splitting associated with each NV crystallographic orientation are labeled. Each $^{15}$NV hyperfine doublet becomes a triplet (with additional sidebands) caused by simultaneous driving of both electron spin transitions.}
        \label{fig:fullODMR}
    \end{figure*}
    \subsection{Vector magnetometry with NV ensembles}\label{vectormagnetometry}
    In diamond, vector magnetometry is enabled by the measurement of the magnetic field along the four possible crystallographic orientations of NV centers denoted by $[111]$, $[\bar{1}\bar{1}1]$, $[\bar{1}1\bar{1}]$, and $[1\bar{1}\bar{1}]$. The four NV orientations can be parameterized by four unit vectors $\vec{o}_i$:
    \begin{ceqn} 
    \begin{align}
    \vec{o}_1 = \frac{1}{\sqrt{3}}(1,1,1),
    \end{align}
    \end{ceqn}
    \begin{ceqn} 
    \begin{align} 
    \vec{o}_2 = \frac{1}{\sqrt{3}}(-1,-1,1),
    \end{align}
    \end{ceqn}
    \begin{ceqn} 
    \begin{align}
    \vec{o}_3 = \frac{1}{\sqrt{3}}(-1,1,-1),
    \end{align}
    \end{ceqn}
    \begin{ceqn} 
    \begin{align}
    \vec{o}_4 = \frac{1}{\sqrt{3}}(1,-1,-1).
    \end{align}
    \end{ceqn}
    With an arbitrarily oriented magnetic field $\vec{B}$, the four Zeeman projections $Z_i\propto \vec{o}_i\cdot\vec{B}$ are measured by measuring the full optically-detected-magnetic-resonance (ODMR) spectrum comprising eight resonances: two of each $m_S=0\leftrightarrow\pm1$ transition for each of four orientations as seen in Fig.~\ref{fig:fullODMR}. 
    
    In this work platform, because of the $^{15}$NV hyperfine-mixed driving, each hyperfine ODMR doublet becomes a triplet with additional sidebands. To account for these extra sidebands, the normalized ODMR spectrum is fit to a multi-dip function:
    \begin{ceqn} 
    \begin{align}
        L(a,b_{jk},g_{jk},Z_j,D,M_j,A_{hyp},f) = a -\sum_{j=1}^{8}\sum_{k=0}^{2} b_{jk} \frac{g_{jk}^2}{(f-(D+M_j+Z_j\pm kA_{hyp}))^2+g_{jk}^2}
        \label{equation:fulldipfits}
    \end{align}
    \end{ceqn}
    where $a$ is the off-resonant vertical offset, $b_{jk}$ is related to depth of each dip, $g_{jk}$ is related to the linewidth of each ODMR dip, $D$ is the zero-field-splitting, $A_{hyp}$ is the diagonalized hyperfine interaction strength (approx 3.05~MHz), $M_j$ is the on-axis strain coefficient, and $Z_j$ is the Zeeman projection along each NV orientation. Note that $Z_j|_{j<4}$~=~$Z_j$ and $Z_j|_{j>4}$~=~$-Z_{5-j}$ and $M_j|_{j>4}$=$M_{(5-j)}$ so that there are four total of each quantity $Z_i$ and $M_i$, one for each NV orientation. Using Eq.~\ref{equation:fulldipfits}, the ODMR spectrum is fit and the four Zeeman projections are extracted (Fig.~\ref{fig:fullODMR}).
    
    In order to transform the Zeeman projections into the lab frame, a linear mapping from $Z_i$ to lab frame is used:
    \begin{ceqn} 
    \begin{align}
    \begin{pmatrix} 
    B_x\\
    B_y\\ 
    B_z
    \end{pmatrix}
    = \frac{1}{\gamma_{NV}}\frac{\sqrt{3}}{8}
    \begin{pmatrix} 
    1 & -1 & -1 & 1\\
    1 & -1 & 1 & -1\\
    1 & 1 & -1 & -1
    \end{pmatrix}
    \begin{pmatrix} 
    Z_1\\ 
    Z_2 \\ 
    Z_3 \\ 
    Z_4 
    \end{pmatrix},
    \label{equation:ZeemantoLab}
    \end{align}
    \end{ceqn}
    where $\gamma_{NV}$=28~MHz/mT is the NV gyromagnetic ratio. This method is used to compute probe magnetic field vector maps such as in Fig.~1d in the main text. The full ODMR spectrum is measured in wide-field, and four Zeeman projection maps (Fig.~\ref{fig:zeemanmaps}) are constructed after fitting the spectrum at each camera pixel. Using Eq.~\ref{equation:ZeemantoLab}, the Zeeman maps (Fig.~\ref{fig:zeemanmaps}) are converted to magnetic field component maps (Fig.~\ref{fig:vectorcomponents}) with respect to lab-frame coordinates. 

    \renewcommand{\thefigure}{S\arabic{figure}}
    \begin{figure*}
        \centering
        \includegraphics[scale=0.3]{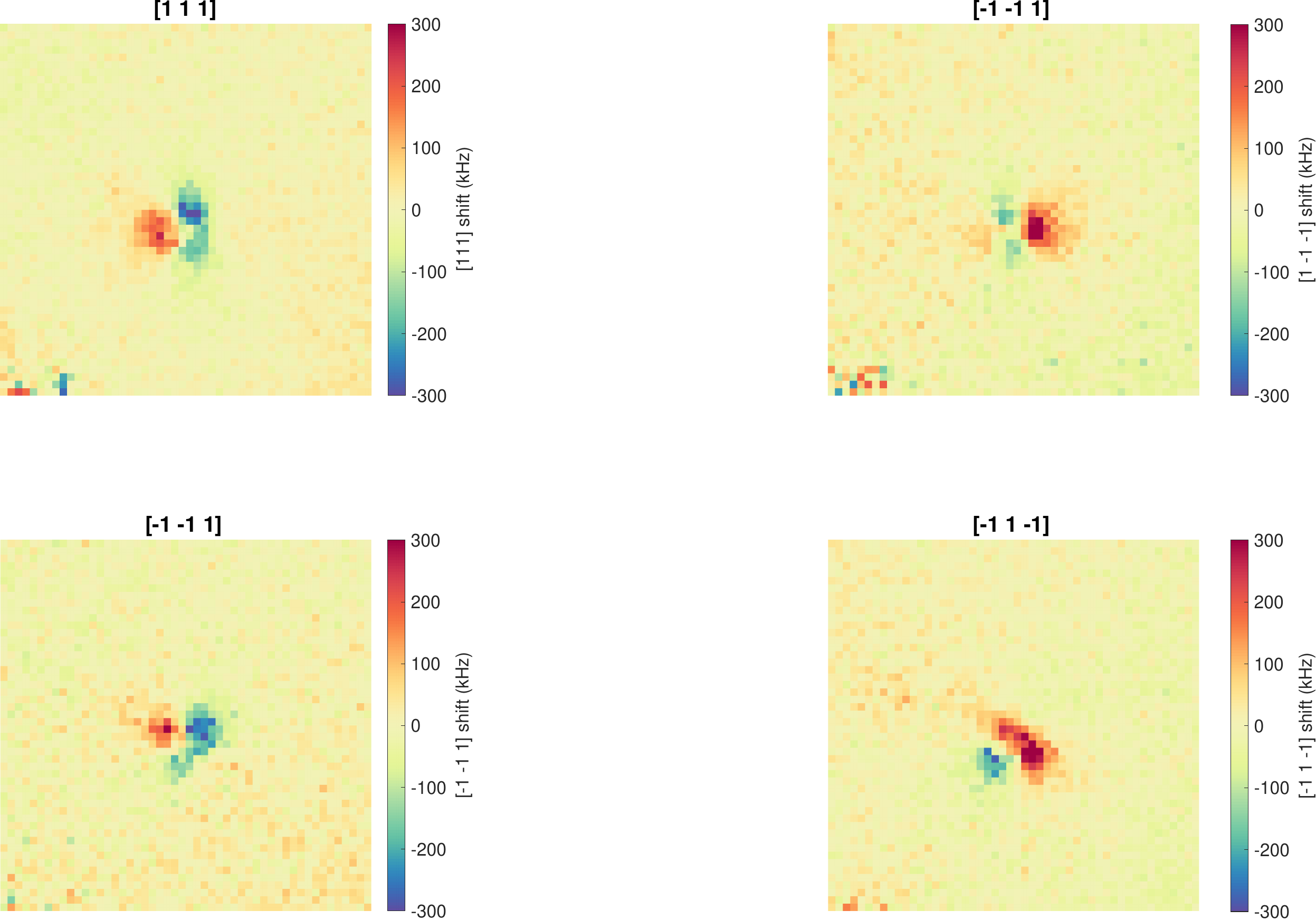}
        \caption{\textbf{Zeeman projection maps.} The four Zeeman projections are measured in wide-field by measuring the full ODMR spectrum at each pixel. Each Zeeman projection is associated with the magnetic field projection along one NV orientation. Image side length is 20~\textmu m.}
        \label{fig:zeemanmaps}
    \end{figure*}

    \renewcommand{\thefigure}{S\arabic{figure}}
    \begin{figure*}
        \centering
        \includegraphics[scale=0.35]{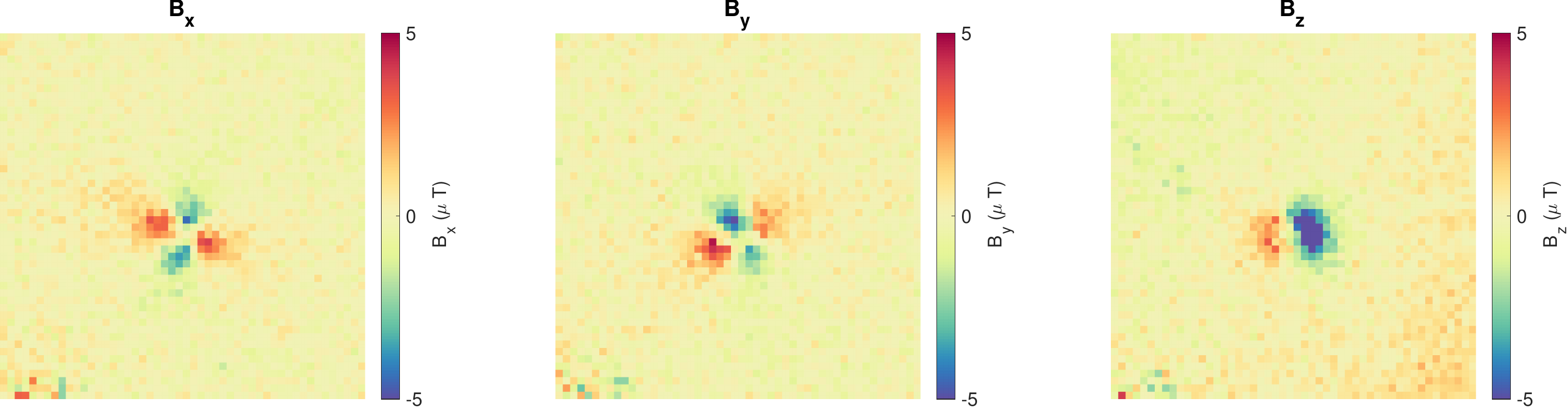}
        \caption{\textbf{Vector component maps.} The four Zeeman projections (Fig.~\ref{fig:zeemanmaps}) are transformed using Eq.~\ref{equation:ZeemantoLab} to vector components. Image side length is 20~\textmu m.}
        \label{fig:vectorcomponents}
    \end{figure*}

    \newpage
    \subsection{Magnetic tweezer applied field}
    In the torque-balance assay, the applied magnetic field exerts a magnetic torque on the DNA-tethered magnetic probe. Additionally, the applied field is asymmetric in order to separate the ODMR from each of the four NV crystallographic orientations. The applied external magnetic field is delivered by three separate magnets. The placement of these magnets relative to the diamond is shown in Fig.~1(b) in the main text. Two SmCo ring magnets are used to exert a homogenous, static, symmetry breaking magnetic field along the NV [111] orientation. Another SmCo disc magnet is used to exert a magnetic tweezer field in the plane of the diamond. The total applied field is changed by rotating the magnetic tweezer in a plane parallel to the diamond. 
    
    \renewcommand{\thefigure}{S\arabic{figure}}
    \begin{figure*}[h]
        \centering
        \includegraphics[scale=0.2]{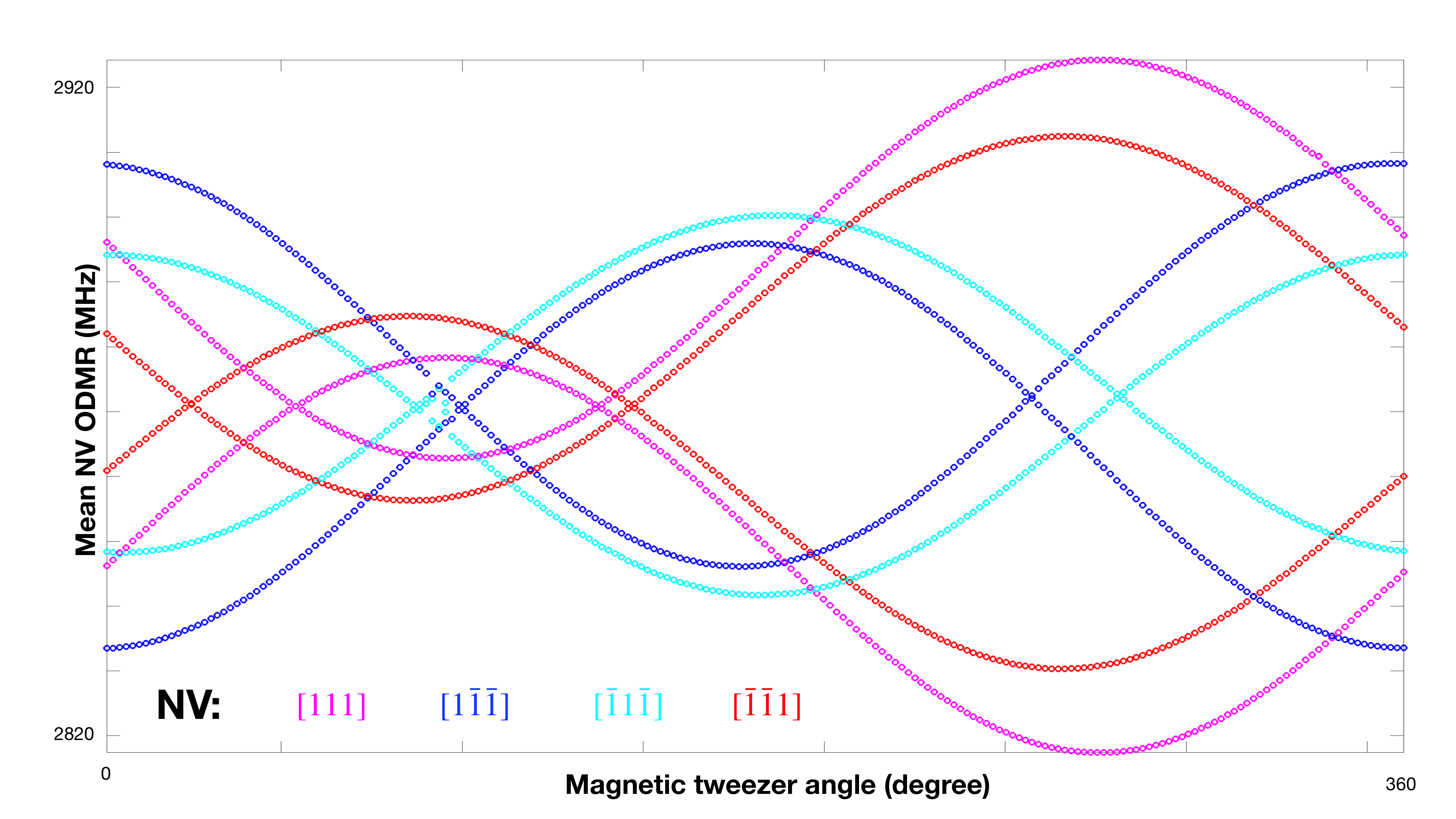}
        \caption{\textbf{ODMR frequencies as a function of magnetic tweezer field angle.} The ODMR associated with each NV crystallographic orientation are color coded. The NV orientations can be identified because the applied field is strongest when the in-plane tweezer field aligns with the static field oriented parallel to the [111] orientation.}
        \label{fig:MTanglesweep}
    \end{figure*}

    \newpage
    \renewcommand{\thefigure}{S\arabic{figure}}
    \begin{figure*}[h!]
        \centering
        \includegraphics[scale=0.35]{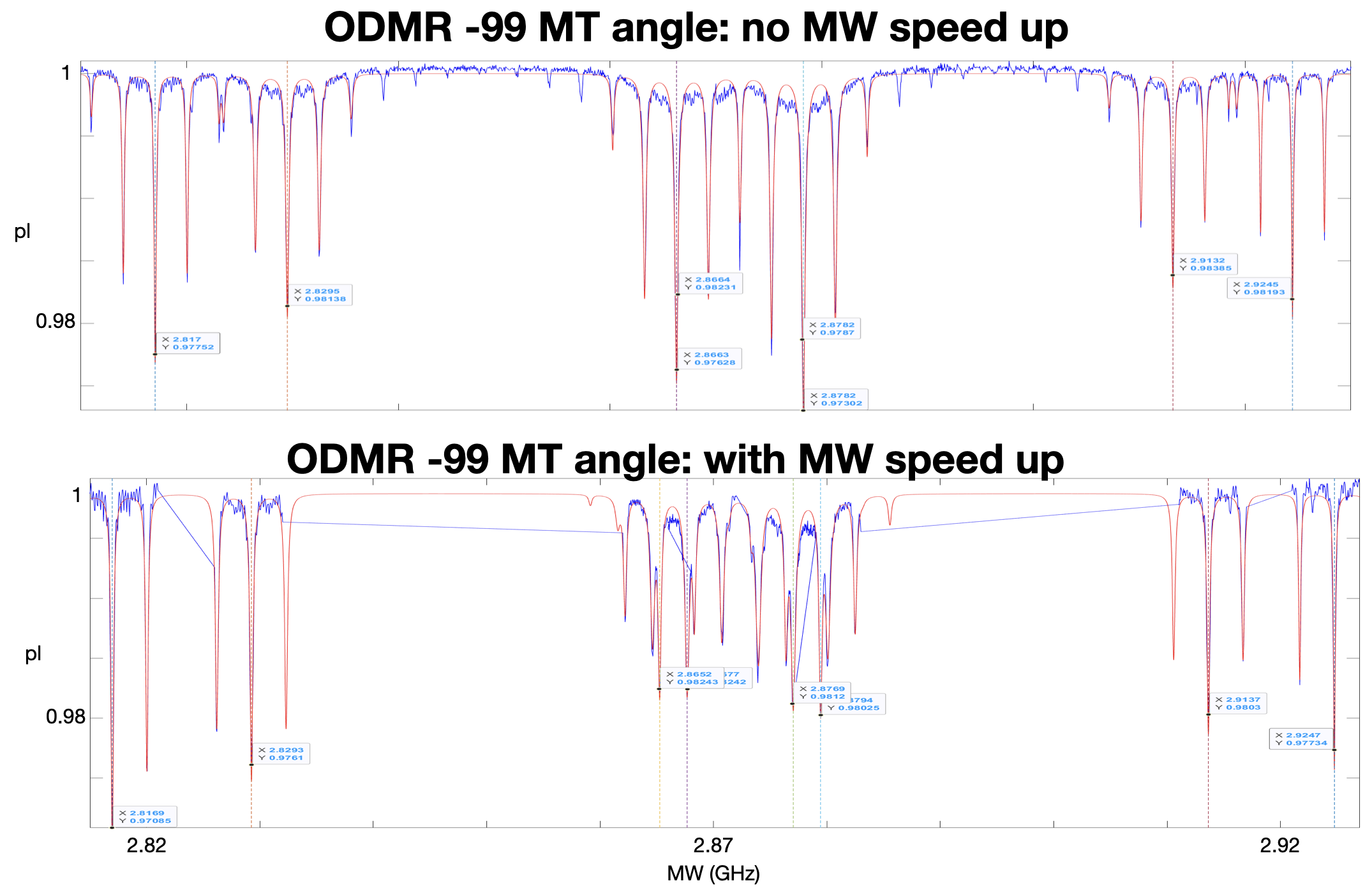}
        \caption{\textbf{ODMR spectrum speed-up comparison.} Top: ODMR spectrum measured by scanning approximately 110~MHz in 50~kHz steps. Bottom: ODMR spectrum measured by only sampling near the main resonance, comprising 24~MHz measured in 50~kHz steps.}
        \label{fig:ODMRcomparison}
    \end{figure*}
    
    \subsection{Applied field calibration to reduce ODMR measurement time}
    In this work, to minimize phototoxicity of the DNA tethers, only a subset MW frequencies are scanned over for a given applied field direction and magnitude. To do this, the full ODMR spectrum due to the total applied field is measured in the absence of a magnetic probe as in Fig.~\ref{fig:MTanglesweep} as a function of magnetic tweezer rotation angle. Then, during the torque-balance experiment, MW frequencies are applied only in a 3~MHz window around each ODMR. In this way, the total data acquisition time is reduced by a factor of approximately 4. An example of this procedure is shown in Fig.~\ref{fig:ODMRcomparison}.

    \newpage
    \renewcommand{\thefigure}{S\arabic{figure}}
    \begin{figure*}
        \centering
        \includegraphics[scale=0.3]{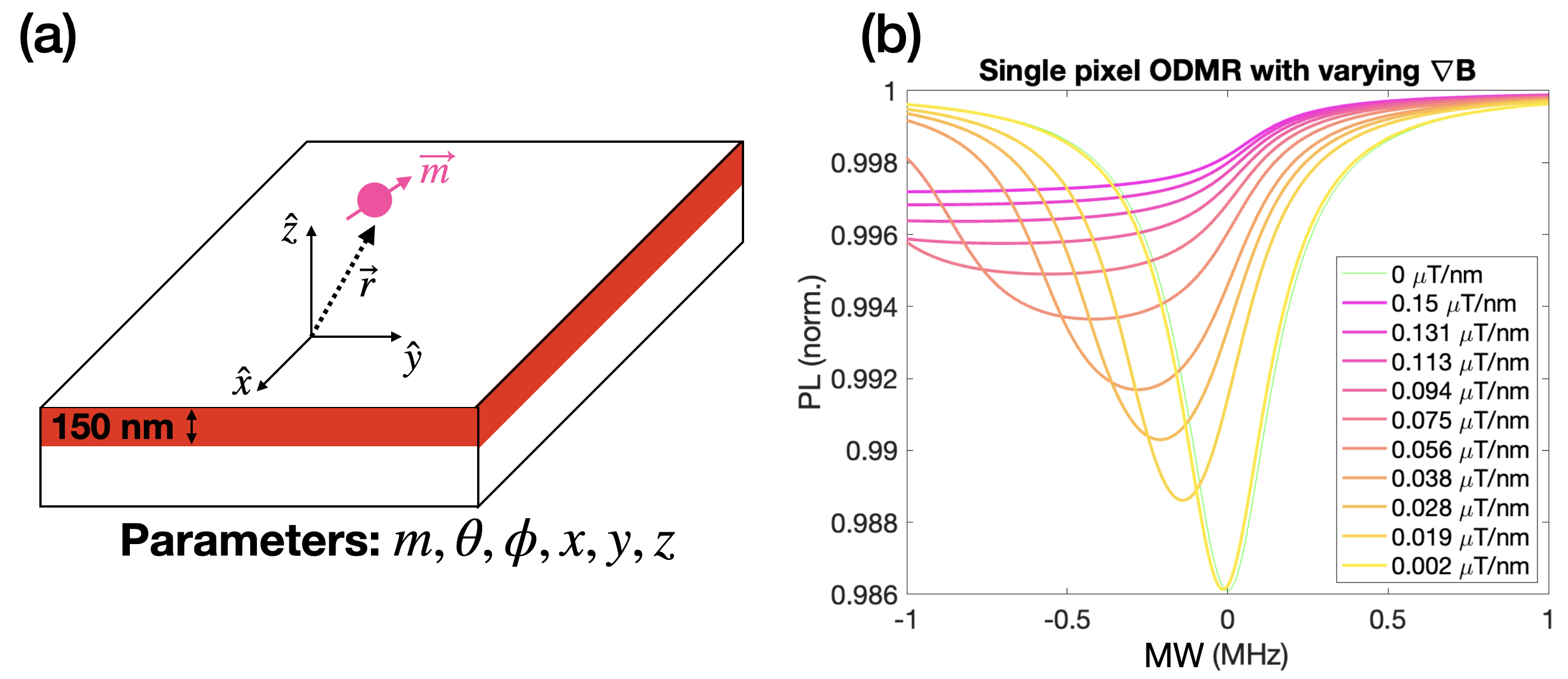}
        \caption{\textbf{Probe field point dipole model and magnetic field gradient induced inhomogenous broadening.}  Schematic of dipole model used to simulate magnetic particle images. Six parameters are used in the model: magnetic dipole moment magnitude, magnetic moment polar and in-plane angles, and three dimensional position vector. (b) Simulation of experimental inhomogenous broadening due to probe-field magnetic field gradients (different colors given by simulated gradient in inset) in a single pixel, limiting the range of magnetic field gradients that can be measured.}
        \label{fig:dipmodelandbgradinhom}
    \end{figure*}
    
    \section{Point dipole magnetic field model used to fit probe orientation}
    
    To extract magnetic probe moment orientation from magPI vector magnetic field images, a six-parameter dipole model is used. A point dipole with magnetic moment $\vec{m}$ is simulated at a position $\vec{r}$ above a diamond sensor:
        \begin{ceqn} 
        \begin{align} 
        \label{equation:dipolefield}
        \vec{B}_{\textrm{dip}}= \frac{\mu_0}{4\pi} ( \frac{3\vec{r}(\vec{m}\cdot\vec{r})}{r^5}-\frac{\vec{m}}{r^3} )
        \end{align}
        \end{ceqn}
        where $\mu_0$ is the vacuum permittivity. The magnetic field is calculated at a layer 75~nm deep in the diamond sensor, with spacing according to average NV density $\rho_{NV}$. The average NV-NV spacing $\langle{d}_{NV}\rangle$ is given by $\rho_{NV} \langle{d}_{NV}\rangle^3=1$. With $10^{16}$~cm$^{-3}$ NV density, $\langle{d}_{NV}\rangle\approx$ 45~nm. The signal at each NV is then Gaussian filtered with $\sigma$ given by the optical resolution of the system, where the optical FWHM is measured to be approximately 500~nm, so $\sigma \approx 212$~nm. Next, the magnetic field signal is binned to the size of individual camera pixels (560~nm). 

    \subsection{Gradient mask}
    \label{gradientmask}
    This magnetic field imaging system is limited by the magnetic gradient due to inhomogenous broadening of NVs within one camera pixel. Multiple NVs contribute to the ODMR signal in a single pixel, so if there is a large magnetic field gradient across the pixel the individual ODMR signals do not destructively sum together, resulting in a dip that is not detectable above the noise as illustrated in Fig.~\ref{fig:dipmodelandbgradinhom}(b). Thus, in the dipole model, after the the gradient of the dipole magnetic field along each NV orientation in a single pixel is calculated, pixels with inhomogenous broadening higher than a threshold value are not used in the fit. 
    
\begin{figure*}
    \begin{center}
     \includegraphics[scale=0.35]{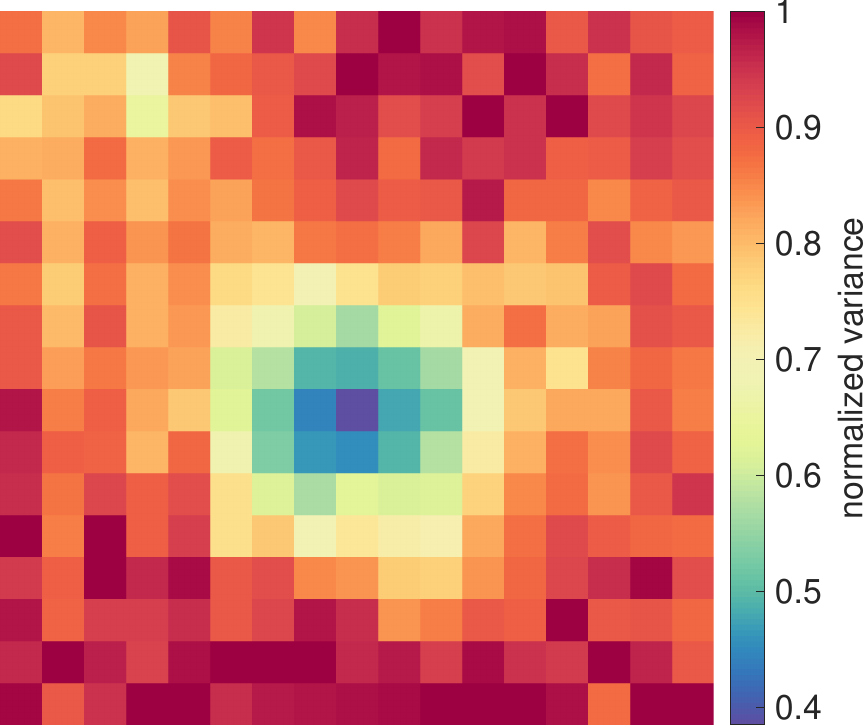}
     \end{center}
    \caption{\textbf{Magnetic field gradient thresholded by examining variance in normalized PL.} Relative variance of normalized PL of the bead region shown in Fig.~\ref{fig:fitwithmask}. The region with low relative variance corresponds to a high magnetic field gradient. Pixels below 0.7 normalized variance are masked.}
    \label{fig:plvariance}
    \end{figure*}

\begin{figure*}
    \begin{center}
     \includegraphics[scale=0.25]{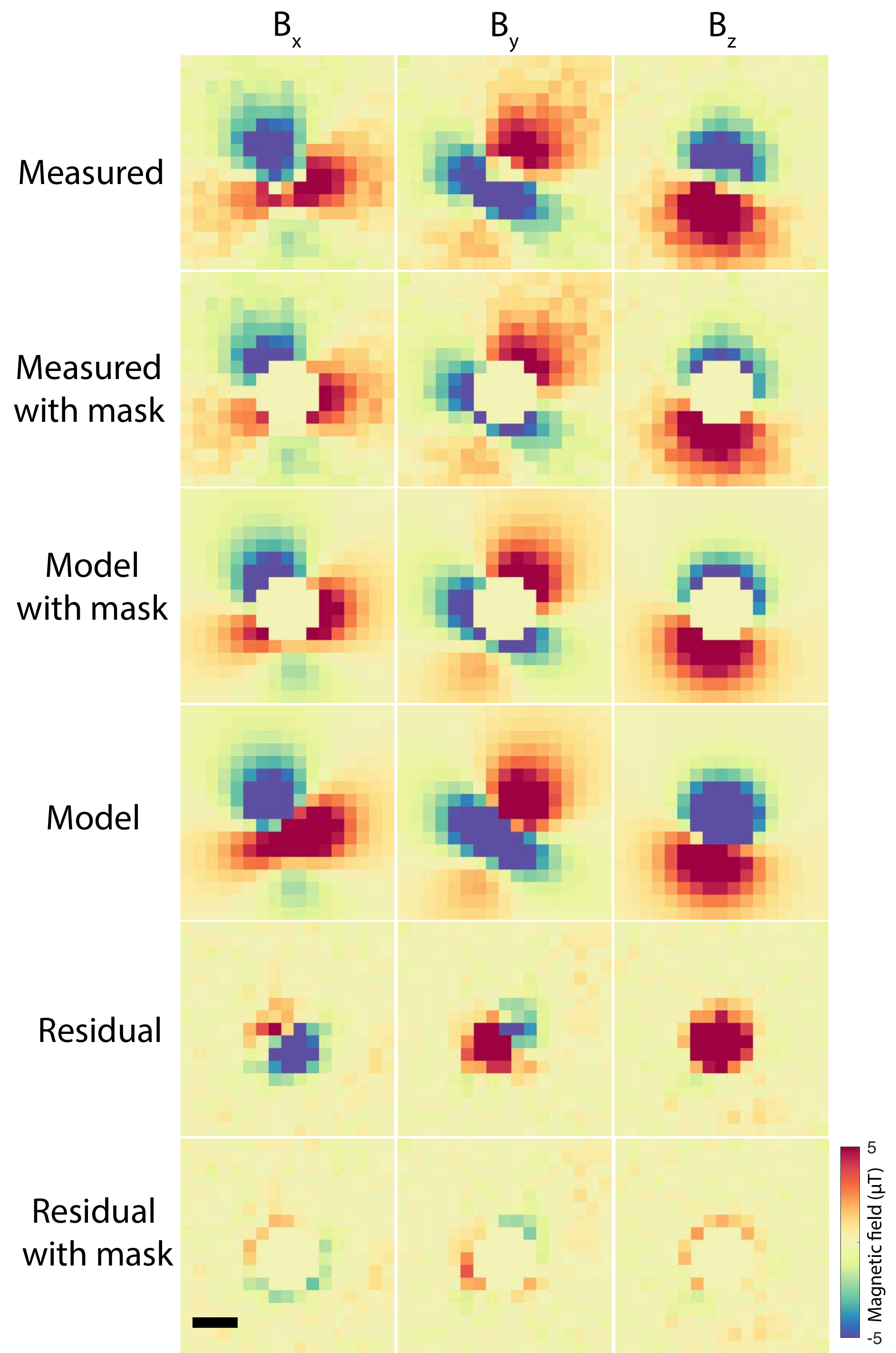}
     \end{center}
    \caption{\textbf{Dipole model with gradient mask fits experimental probe field images.} Experimental data is fit using a dipole model with experimental gradient mask. Residual is the difference between measured and model, and residual with mask is the difference between measured with mask and model with mask. Scale bar is 2~\textmu m.}
    \label{fig:fitwithmask}
    \end{figure*}

    The amount of inhomogenous broadening can be quantified by examining the variance in normalized photoluminescence (PL). During the ODMR measurement, PL is collected as a function of MW frequency at each pixel. The variance in the distribution of these measured PL values at each pixel scales with the amount of inhomogenous broadening. Pixels with large magnetic field gradients and thus large inhomogenous broadening show low variance in PL, and pixels with small magnetic field gradients show a high variance in PL as shown in Fig.~\ref{fig:plvariance}. To mask large gradients, a threshold value of relative variance is chosen, and pixels below this threshold are not used in the fit. Probe field fitting with gradient masking is shown in Fig.~\ref{fig:fitwithmask}. In the figure, a series of vector magnetic field images are plotted: the first row shows the measured vector component images for a probe orientation. The second row shows the measured vector component images with gradient mask applied. The third row shows the fitted model probe images with gradient mask applied. The fourth row shows the fitted model probe images without gradient mask applied. The fifth row shows the residual between measured and model probe images. The sixth row shows the residual between measured with mask and model with mask probe images. 
    
    \subsection{Error analysis}
    Error on the data in the main text (Figs.~3 and 4) is given by the error in the probe dipole model fit parameters. Error in each fit parameter $\alpha_j$ is given by 
    \begin{ceqn} 
    \begin{align} 
        \delta\alpha_j = \sqrt{\textrm{Cov}_{jj}}*\textrm{var}(\textrm{R})
    \end{align}
    \end{ceqn}
    where $\textrm{Cov}_{jj}$ is the diagonal element of the covariance matrix corresponding to parameter $j$ and $\textrm{R}$ is the residual between fit images and model images with gradient masking. This method is used to determine the error in probe in-plane angle, probe polar angle, and probe x and y positions. In the dipole model, probe moment and probe height are fixed, and the error in moment and probe height are discussed in Section~\ref{momentprecision}.

    After determining the error of the fit parameters, the error in the measured quantities is determined using standard error propagation, for example:
    \begin{ceqn} 
    \begin{align} \label{equation:torquezerror}
    \delta \tau_z = (|m|\cos\theta_\textrm{m}(B_y\cos\phi_\textrm{m}-B_x\sin\phi_\textrm{m}))\delta\theta_\textrm{m}\\
    +(|m|(B_y\sin\theta_\textrm{m}\sin\phi_\textrm{m}+B_x\sin\theta_\textrm{m}\cos\phi_\textrm{m}))\delta\phi_\textrm{m}\\
    +(|m|(B_y\sin\theta_\textrm{m}\cos\phi_\textrm{m}-B_x\sin\theta_\textrm{m}\sin\phi_\textrm{m}))\delta m
    \end{align}
    \end{ceqn}
    
    Because the millitesla applied field is determined by averaging the signal from several thousands of pixels each with approximately 100~nT/Hz$^{1/2}$ sensitivity~\cite{Kazi2021-dy}, the errors in the applied field are found to be negligible. 

    \begin{figure*}
    \begin{center}
     \includegraphics[scale=0.28]{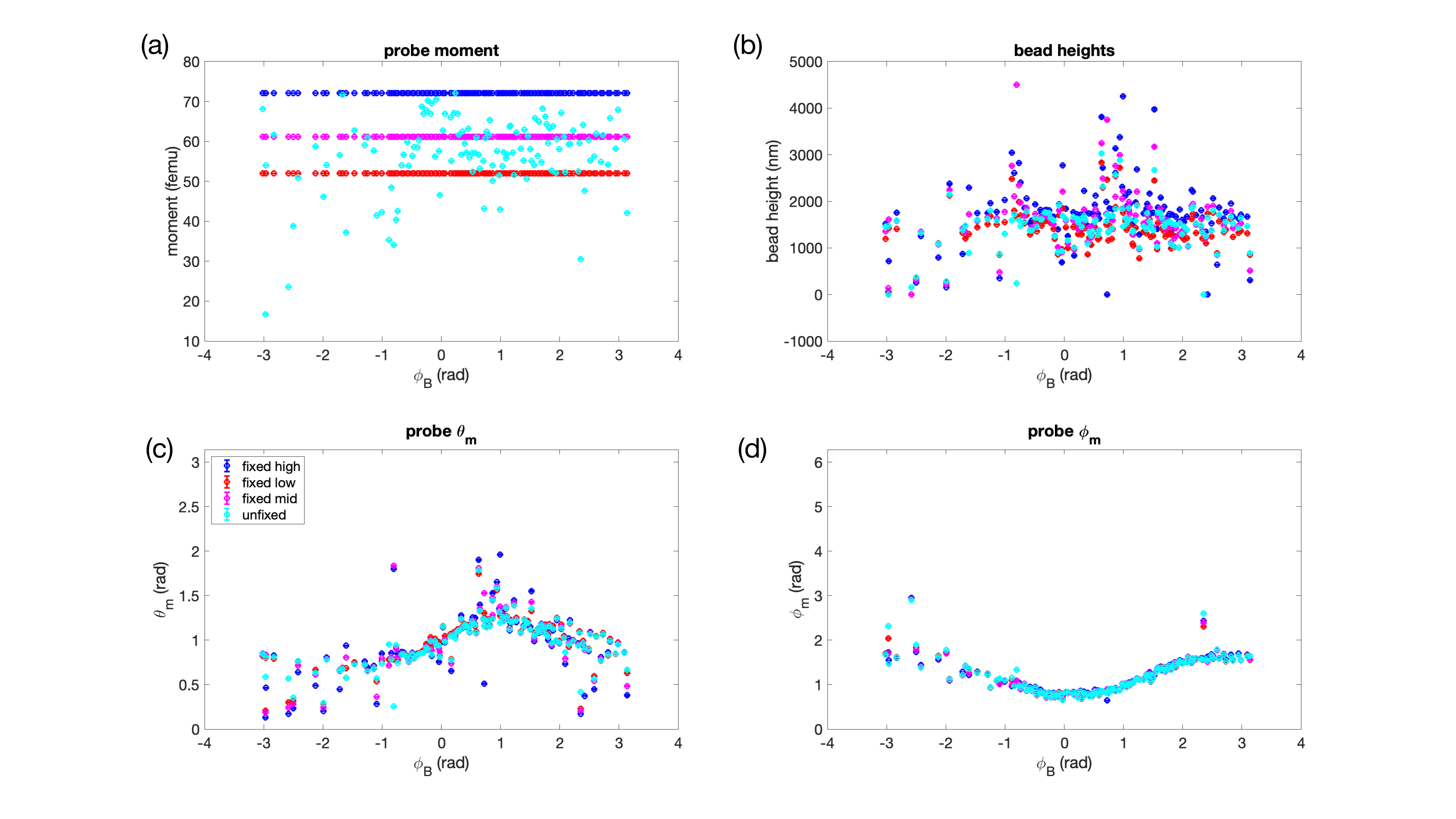}
     \end{center}
    \caption{\textbf{Moment precision determined empirically.} Four different guesses of moment magnitude (fixed high, mid and low, and unfixed) are used to fit the same data sets to determine the precision of probe magnetic moment fitting. (a) Fitted probe moment as a function of applied field in-plane angle $\phi_\textrm{B}$. (b) Fitted probe height as a function of applied field angle. (c) Fitted probe polar angle $\theta_\textrm{m}$ as a function of applied field in-plane angle $\phi_\textrm{B}$. (d) Probe in-plane angle $\phi_\textrm{m}$ as a function of applied field in-plane angle $\phi_\textrm{B}$. From the data, the spread in moment magnitude is found to weakly affect fitted probe orientation and probe height.}
    \label{fig:momentspread}
    \end{figure*}

        \subsubsection{Moment magnitude and probe height precision} \label{momentprecision}
        As seen in Eq.~\ref{equation:dipolefield}, the dipole field depends on moment magnitude, probe position and probe orientation. In order to extract precision using least-squares fitting, the covariance matrix element of the parameter is used, requiring the parameter to float in the fitting process. However, in subsequent applied field positions, the probe orientation can change but the probe moment magnitude is fixed. Thus, in this work, to estimate the probe moment magnitude precision, many images are fit with the moment allowed to vary. Then, the distribution of fitted moment magnitudes provides the mean moment and variance of the moment magnitude as seen in Fig.~\ref{fig:momentspread}. For most data sets, the standard deviation in probe moment magnitude is found to be approximately 10\%. 
        
        \begin{figure*}
        \begin{center}
         \includegraphics[scale=0.5]{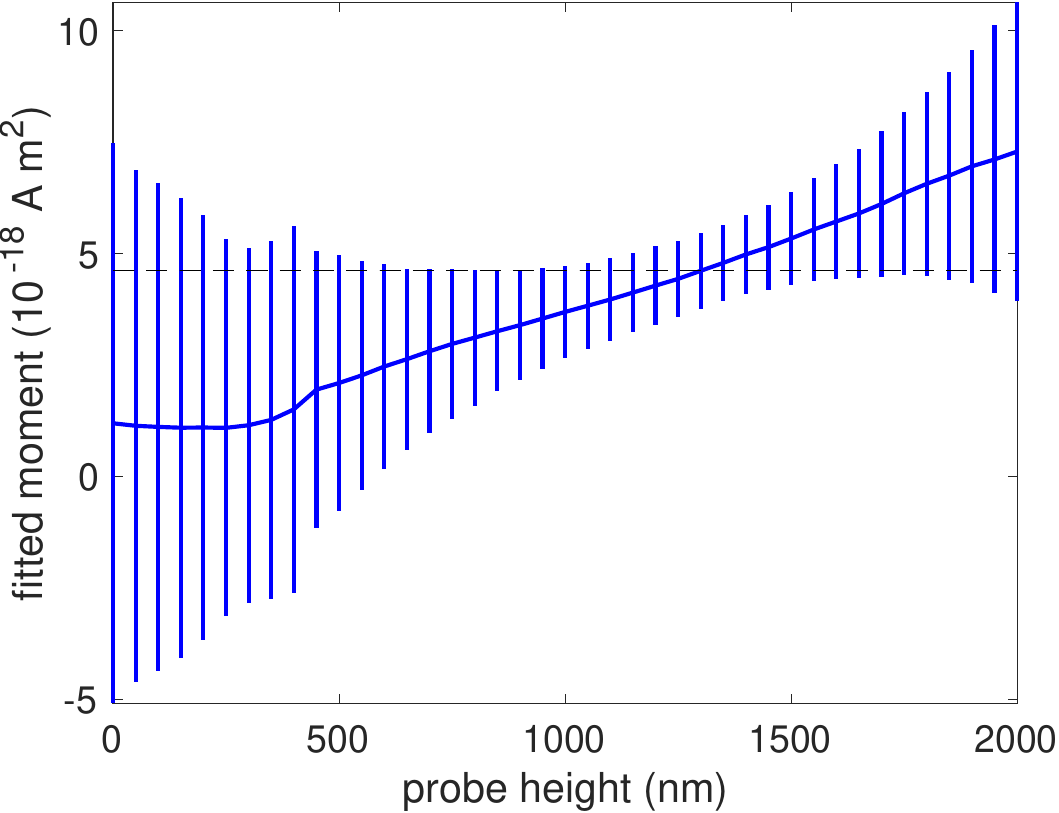}
         \end{center}
        \caption{\textbf{Change in moment magnitude with fixed probe height.} Probe moment is fit as a function of fixed probe height for the dataset shown in Fig.~\ref{fig:fitwithmask}. The probe moment value associated with the best fit (lowest residual) is shown by the horizontal dashed line.}
        \label{fig:momentandheight}
        \end{figure*}
        
        We find the dipole model probe height fit parameter determined by least-squares fitting to be larger than expected. For example, in the data shown in Fig.~\ref{fig:fitwithmask} showing the probe field for a 200\;nm DNA tether, the fitted probe height above the diamond is $\approx$ 1200\;nm, much larger than the length of the DNA.
        
        While the probe height does not directly enter into the torque calculation, for a given height there is a different best fit probe moment - which is crucial for determining magnitude of applied torque on the DNA tethers. We illustrate the relationship between fitted probe moment and probe height in Fig.~\ref{fig:momentandheight}, where the probe moment is fit for the dataset shown in Fig.~\ref{fig:fitwithmask} with fixed probe height for each fit. The horizontal dashed line corresponds to the fitted probe value that has the lowest residual and thus best fits the data. 
        
        Fig.~\ref{fig:momentandheight} shows that the best fit probe moment is within the probe moment uncertainty for a wide range of probe heights, and that we may be overestimating our probe moment values by a factor of 3. The cause of obtaining large probe heights is unknown, but may be caused by the imaging platform's inability to detect magnetic field for pixels directly below the bead due to the strong magnetic field gradients (Sec.~\ref{gradientmask}), which may be ameliorated by using super-resolution magnetic imaging~\cite{Mosavian2023-sc}. Developing the magnetic dipole model and understanding the relationship between probe height and probe moment magnitude is key to advancing the torque balance technology further will be the subject of future work.

    \section{Additional torque-balance measurements on DNA tethers}
    
    \begin{figure*}
    \begin{center}
     \includegraphics[scale=0.9]{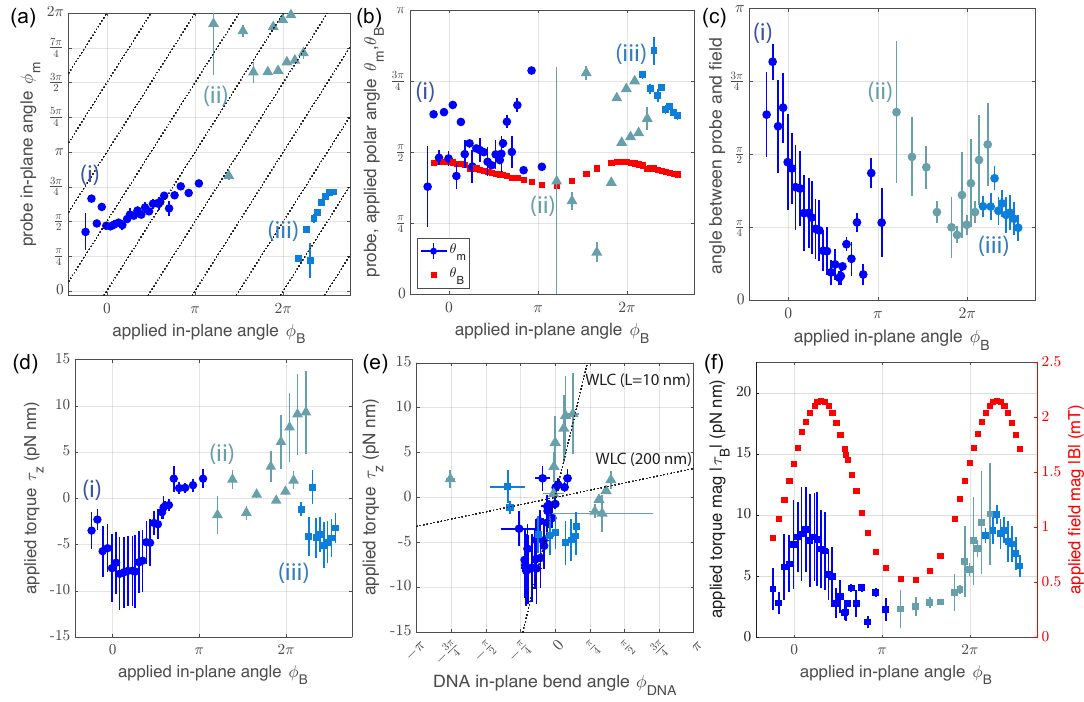}
     \end{center}
    \caption{\textbf{Response of dipole to applied field for additional 200\;nm DNA molecule.} (a) Probe moment in-plane angle $\phi_\textrm{m}$ trajectory as a function of applied field in-plane angle $\phi_\textrm{B}$ for a 200\;nm DNA tether. Dashed lines with unit slope are plotted and represent the response of a torsion-free DNA tether. Deviations from dashed lines indicate the DNA is exerting a torque on the magnetic probe. Three trajectories are observed in which the DNA tether undergoes a distinct response (different colors). (b) Probe moment polar angle $\theta_\textrm{m}$ and applied field polar angle $\theta_\textrm{B}$ as a function of applied field in-plane angle $\phi_\textrm{B}$. Again, deviations between the probe direction and the applied field direction during the three trajectories indicate the DNA is exerting a torque on the magnetic probe. (c) Angle
    between probe moment vector and applied field vectors as a function of applied field in-plane angle $\phi_\textrm{B}$. (d) Out-of-plane magnetic torque $\tau_z$ as a function of applied applied field in-plane angle $\phi_\textrm{B}$. (e) Out-of-plane magnetic torque $\tau_z$ as a function of DNA in-plane bend angle $\phi_{\textrm{DNA}}$. (d) Magnitude of applied torque $|\tau_\textrm{B}|$ and applied field magnitude $|B|$ as a function of applied field in-plane angle $\phi_\textrm{B}$. For this DNA tether, fitted probe moment magnitude m=4.8e-18~Am$^2$.}
    \label{fig:600bp12}
    \end{figure*}
    
    In this section, several other individual DNA molecule bend torque responses are given to demonstrate the assay's ability to measure the behavior of different types of DNA molecules: additional high-stiffness tethers, and a torsion-free DNA tether.

    \subsection{Additional high stiffness DNA molecules}
    The torque-response of two additional nominally 200\;nm tethers are given in this section. In Fig.~\ref{fig:600bp12}(a)\&(b), the probe in-plane and polar angles as a function of applied field in-plane angle are shown. As in Sec.IV of the main text, the probe examined here undergoes three trajectories that show a distinct response (different colors). In trajectory (i), the probe in-plane angle orients with applied field, but with a lower slope than the torsion-free response (dashed lines), after which the probe angle changes largely (ii). In Fig.~\ref{fig:600bp12}(c), the angle between probe orientation and applied field direction is shown. This angle scales with the magnitude of applied torque on the magnetic probe, but in the low applied torque regime (ii) the probe experiences a torque by only the DNA molecule and the angle between probe and applied field changes largely. The out-of-plane torque $\tau_z$ as a function of applied field angle is shown in Fig.~\ref{fig:600bp12}(d), showing a distinct response between the three trajectories. The bend-torque response for this DNA tether is shown in Fig.~\ref{fig:600bp12}(e), with WLC prediction overlaid. This DNA tether appears to exhibit a linear bend response with a higher slope than the WLC prediction, and fitting the slope of the trajectory to a line extracts $L\approx$10\;nm with $L_p$=50\;nm. At the end of trajectory (iii) the DNA tether signal was lost.

    \newpage 
    
    \begin{figure*}
    \begin{center}
     \includegraphics[scale=0.9]{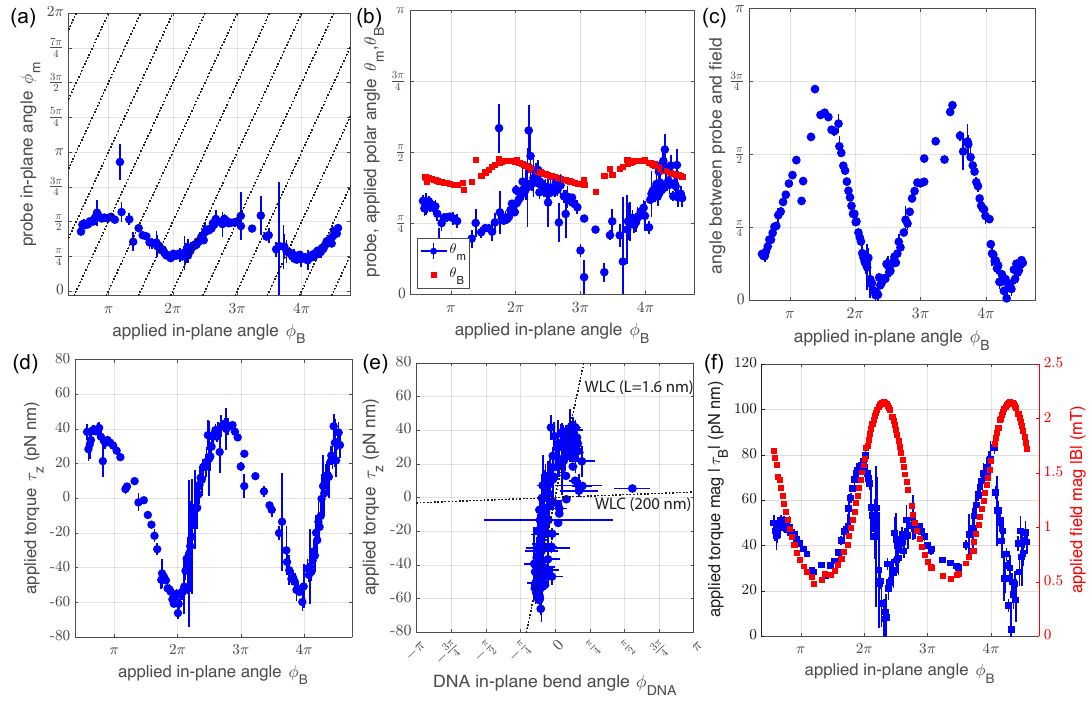}
     \end{center}
    \caption{\textbf{Response of dipole to applied field for high-stiffness DNA tether.} (a) Probe moment in-plane angle $\phi_\textrm{m}$  trajectory as a function of applied field in-plane angle $\phi_\textrm{B}$ for an unknown length DNA tether. Dashed lines with unit slope are plotted and represent the response of a torsion-free DNA tether. Deviations from dashed lines indicate the DNA is exerting a torque on the magnetic probe. (b) Probe moment polar angle $\theta_\textrm{m}$ and applied field polar angle $\theta_\textrm{B}$ as a function of applied field in-plane angle $\phi_\textrm{B}$. Again, deviations between the probe direction and the applied field direction during the three trajectories indicate the DNA is exerting a torque on the magnetic probe. (c) Angle
    between probe moment vector and applied field vectors as a function of applied field in-plane angle $\phi_\textrm{B}$. (e) Out-of-plane magnetic torque $\tau_z$ as a function of applied field in-plane angle $\phi_\textrm{B}$. (e) Out-of-plane magnetic torque $\tau_z$ as a function of DNA in-plane bend angle $\phi_{\textrm{DNA}}$. (d) Magnitude of applied torque $|\tau_\textrm{B}|$ and applied field magnitude $|B|$ as a function of applied field in-plane angle $\phi_\textrm{B}$.  For this DNA tether, fitted probe moment magnitude m=61e-18~Am$^2$.}
    \label{fig:extrastifftether}
    \end{figure*}
    
    The measured response of the torque balance for another high-stiffness DNA molecule of nominally 200\;nm length is shown in Fig.~\ref{fig:extrastifftether}. This tether was attached to a large clump of ferromagnetic material with total fitted magnetic moment magnitude m=61e-18~Am$^2$. The tether may be a shunted tether in which the surface-DNA and DNA-probe linker oligos are directly ligated together, and may represent the extremely short tether limit. In Fig.~\ref{fig:extrastifftether}(a), the probe in-plane angle is measured as a function of applied field in-plane angle and in Fig.~\ref{fig:extrastifftether}(b) the probe and applied field polar angles are measured as a function of applied field in-plane angle. For this tether, the probe orientation is only slightly perturbed in a sinusoidal fashion by the applied field and can be understood as a nutation of probe orientation about a ``home" orientation caused by the influence of the applied field. In Fig.~\ref{fig:extrastifftether}(c), the angle between probe orientation and applied field direction is shown, consistent with the nutation interpretation. Similarly, the out-of-plane torque $\tau_z$ as a function of applied field angle shown in Fig.~\ref{fig:extrastifftether}(d) oscillates as a function of applied field orientation. The bend-torque response for this DNA tether is shown in Fig.~\ref{fig:extrastifftether}(e), with WLC prediction ($L=$200\;nm, $L_p$=50\;nm) overlaid. The DNA tether experiences a significantly larger torque than would be expected for a 200\;nm molecule. Fitting the slope of the trajectory to a line extracts $L\approx$1.6\;nm with $L_p$=50\;nm. 
    The total applied torque and applied field magnitude in Fig.~\ref{fig:extrastifftether}(f) shows that the DNA tether buckles for total applied torque around 75~pN~nm, which could provide additional input about the biophysical structure of the tether. This experiment shows that the torque-balance assay can examine the bend energy of a high stiffness DNA molecule, but the structure of the DNA tether must be validated by an orthogonal imaging modality in order to fully interpret the bend energy response.

    \begin{figure*}
    \begin{center}
     \includegraphics[scale=0.9]{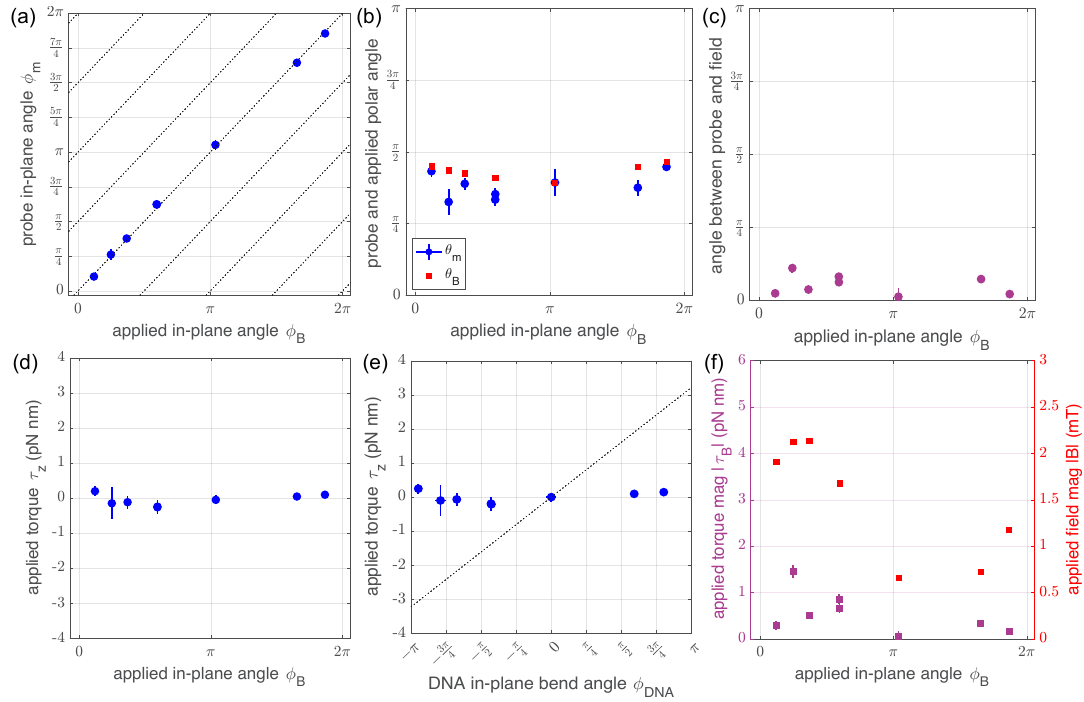}
     \end{center}
    \caption{\textbf{Response of dipole to applied field for 200\;nm torsion-free DNA molecule.} (a) Probe moment in-plane angle $\phi_\textrm{m}$ trajectory as a
    function of applied field in-plane angle $\phi_\textrm{B}$. (b) Probe moment polar angle $\theta_\textrm{m}$ and applied field polar angle $\theta_\textrm{B}$ as a function of applied field in-plane angle $\phi_\textrm{B}$. (c) Angle between probe moment vector and applied field vectors as a function of applied field in-plane angle $\phi_\textrm{B}$. (d) Out-of-plane applied torque $\tau_z$ as a function of applied field in-plane angle $\phi_\textrm{B}$. (e) Out-of-plane magnetic torque $\tau_{z}$ as a function of DNA in-plane bend angle $\phi_{\textrm{DNA}}$. In this case, the DNA ``home" orientation is given by $\langle\phi_\textrm{m}\rangle\approx\pi$. Dashed line is the WLC prediction with $L_p$=50\;nm and $L$=200\;nm. (f) Magnitude of applied torque $|\tau_\textrm{B}|$ and applied field magnitude $|B|$ as a function of applied field in-plane angle $\phi_\textrm{B}$. For this DNA tether, fitted probe moment magnitude m=2.4e-18~Am$^2$.}
    \label{fig:torsionfreetorquebalance}
    \end{figure*}
    
    \subsection{Torsion-free DNA molecule}
    As discussed in Sec.IV of the main text, the torsion-free DNA tether provides important experimental tests. If the torsion-free tether is found to exert a bend torque on the magnetic probe, there is a systematic error in DNA-diamond tethering. The bend-torque response of a torsion-free DNA tether is given in Fig.~\ref{fig:torsionfreetorquebalance}. The probe moment is always mostly oriented along the applied field direction because the DNA is unable to exert a torque on the probe, so the probe in-plane angle $\phi_\textrm{m}\approx\phi_\textrm{B}$ and $\theta_\textrm{m}\approx\theta_\textrm{B}$ as shown in Fig.~\ref{fig:torsionfreetorquebalance}(a)\&(b), while the angle between probe and applied field $\approx0$ shown in Fig.~\ref{fig:torsionfreetorquebalance}(c). For each of these angles, misalignment between probe orientation and applied field direction may be caused by thermal fluctuations of probe moment orientation. The torque applied by the magnetic field on the probe is also close to zero as seen in Fig.~\ref{fig:torsionfreetorquebalance}(d). For the torsion-free tether, there is no well defined home orientation, so the DNA bend angle is ill-defined: for this data the home orientation is arbitrarily chosen as the mid-point of the trajectory $\langle\phi_\textrm{m}\rangle\approx\pi$ [Fig.~\ref{fig:torsionfreetorquebalance}(e)]. Finally, the total applied torque compared to applied field magnitude is shown in Fig.~\ref{fig:torsionfreetorquebalance}(f). The total applied torque is near-zero, and any non-zero values may be caused thermal fluctuations of the probe moment because the thermal energy of a 200\;nm DNA tether at room temperature is approximately 1~pN~nm as discussed in Sec.IIB of the main text. After the data shown was taken, the magnetic probe signal was lost.

\bibliography{supplement}